\documentclass[journal]{IEEEtran}
\usepackage{amsmath,amsfonts}
\usepackage[linesnumbered,ruled]{algorithm2e}
\usepackage{algorithmic}
\usepackage{setspace}
\usepackage{array}
\usepackage[caption=false,font=normalsize,labelfont=sf,textfont=sf]{subfig}
\usepackage{textcomp}
\usepackage{stfloats}
\usepackage{url}
\usepackage{verbatim}
\usepackage{graphicx}
\usepackage{bbm}
\usepackage{latexsym}
\usepackage{cite}
\usepackage{amssymb}
\usepackage{color}
\usepackage{multirow}
\usepackage{amsthm}
\usepackage{booktabs}

\newtheorem{myrmk}{Remark}

\usepackage{xcolor}

\definecolor{ForestGreen}{RGB}{34,139,34}
\newcommand{\ADDED}[1]{#1}

\begin{document}

\title{Online Graph \ADDED{Topology} Learning via\\Time-Vertex Adaptive Filters: From Theory to Cardiac Fibrillation}%

\author{Alexander Jenkins$^{1, 2}$\thanks{$^1$ Department of Electrical and Electronic Engineering, Imperial College London, London, SW7 2AZ, United Kingdom.}\thanks{$^2$ National Heart and Lung Institute, Imperial College London, London, United Kingdom.}, Thiernithi Variddhisai$^{1}$, Ahmed El-Medany$^2$, \\ Fu Siong Ng$^2$ and Danilo Mandic$^{1}$,~\IEEEmembership{Fellow,~IEEE}

\thanks{Correspondence to: Alexander Jenkins $<$a.jenkins21@imperial.ac.uk$>$.}%
\thanks{An earlier version of the core algorithm was presented in preprint form \cite{variddhisai2020methods}. The current manuscript represents a comprehensive advancement with extended theoretical analysis, synthetic experiments, and a real-world case study on cardiac fibrillation.}
\thanks{Code available: \protect\url{https://github.com/jenkins-alex/AdaCGP}}}

\markboth{Journal of \LaTeX\ Class Files,~Vol.~14, No.~8, August~2021}%
{Shell \MakeLowercase{\textit{et al.}}: A Sample Article Using IEEEtran.cls for IEEE Journals}

\maketitle

\begin{abstract}
Graph Signal Processing (GSP) provides a powerful framework for analysing complex, interconnected systems by modelling data as signals on graphs. While recent advances have enabled graph topology learning from observed signals, existing methods often struggle with time-varying systems and real-time applications. To address this gap, we introduce AdaCGP, a sparsity-aware adaptive algorithm for dynamic graph topology estimation from multivariate time series. AdaCGP estimates the Graph Shift Operator (GSO) through recursive update formulae designed to address sparsity, shift-invariance, and bias. Through comprehensive simulations, we demonstrate that AdaCGP consistently outperforms multiple baselines across diverse graph topologies, achieving improvements exceeding 83\% in GSO estimation compared to state-of-the-art methods while maintaining favourable computational scaling properties. Our variable splitting approach enables reliable identification of causal connections with near-zero false alarm rates and minimal missed edges. Applied to cardiac fibrillation recordings, AdaCGP tracks dynamic changes in propagation patterns more effectively than established methods like Granger causality, capturing temporal variations in graph topology that static approaches miss. The algorithm successfully identifies stability characteristics in conduction patterns that may maintain arrhythmias, demonstrating potential for clinical applications in diagnosis and treatment of complex biomedical systems.
\end{abstract}

\begin{IEEEkeywords}
Graph topology estimation, functional connectivity, adaptive graph signal processing, graph shift operator, time-vertex stochastic process, multivariate statistical models, cardiac fibrillation.
\end{IEEEkeywords}

\section{Introduction}
\IEEEPARstart{I}{n} an era of unprecedented data generation, the challenge of understanding complex, time-varying systems has become increasingly critical. Much of this data arrives as simultaneous, long-running time series from diverse sources including financial markets, environmental monitoring stations \cite{yi2015survey}, energy grid networks \cite{gungor2010opportunities}, and biomedical sensors \cite{rudy2017noninvasive}. While often appearing unstructured, these datasets frequently encode underlying spatio-temporal relationships that can reveal system dynamics and topology.

Networks/graphs provide a powerful low-dimensional framework for representing relationships between data sources \cite{mandicp1}, with nodes representing sensors and edges capturing their dependencies. When the graph topology (the GSO) is known, various tools from graph signal processing (GSP) theory can analyse signals residing on the graph \cite{mandicp2, mandicp3}. However, in many real-world scenarios, the underlying graph topology is unknown and must be estimated from observed data, often in an online/adaptive manner as new measurements become available.

\ADDED{This fundamental challenge, known as graph topology estimation, has attracted significant attention in recent years, particularly in clinical medicine, where graph topology estimates may reflect functional connectivity in biomedical systems.} For instance, in complex cardiac arrhythmias such as Ventricular Fibrillation (VF) and Atrial Fibrillation (AF), chaotic electrical propagation manifests as patient-specific `electrophenotypes' \cite{sau2019optimum, ng2020toward}. The ability to capture these electrophenotypes and assess their stability holds crucial diagnostic value, yet current treatments remain ineffective due to our inability to characterise these patterns, leading to poor outcomes and high recurrence rates \cite{sau2019optimum, scherr2015five}. Recent work has applied graph topology estimation to identify fibrillation electrophenotypes \cite{handa2020granger}, but these offline/batch approaches lack the ability to track evolving network dynamics and measure the electrophenotype's stability.

\ADDED{Beyond cardiac applications, many neurological and psychiatric conditions, including epilepsy, Alzheimer's disease, and schizophrenia, have been increasingly recognised as disorders of dysregulated network connectivity \cite{fornito2015connectomics}. Graph-based functional connectivity characterises the interactions between brain regions in both health and disease \cite{fornito2015connectomics}. Conventional approaches such as Pearson's correlation provide a simplistic representation of functional interactions \cite{rubinov2010complex}, while precision matrix methods offer a more refined modelling of indirect influences \cite{marrelec2006partial, graphLasso}. While dynamic functional connectivity techniques, such as employing batch methods in sliding windows, have improved the capture of temporal variations \cite{preti2017dynamic, vidaurre2017brain}, truly adaptive approaches for the task remain in their infancy.}

\ADDED{Alongside neurological and cardiac applications, modelling complex drug-disease interactions represents another critical domain where graph topology estimation proves invaluable. Traditional methods for analysing these interactions often rely on pairwise statistical associations, which fail to capture the underlying causal mechanisms and network-level dependencies that drive disease progression and treatment response \cite{zitnik2018modeling}. Graph-based models provide an effective alternative by representing biomedical systems as structured networks, where nodes correspond to biological entities (genes, proteins, drugs, physiological signals) and edges encode their functional relationships \cite{barabasi2011network}, while causal inference techniques, such as those popularised by Granger \cite{granger1969investigating}, allow the identification of directed interactions by testing whether past observations of one signal improve the prediction of another \cite{barnett2014mvgc}.}

Real-world applications have driven the development of diverse methodological approaches to graph topology estimation, spanning multiple analytical frameworks. Traditional approaches to graph topology estimation include methods based on sparse inverse covariance estimation \cite{graphLasso}, and GSP-based signal smoothness \cite{smoothGraphLearn, dong2016Learning} and diffusion \cite{diffusionNoisySignalsLearning, heatDiffusionGraphs, SegarraSpectralTemplates}. While promising, these methods typically focus on learning variants of the Laplacian or symmetric matrices, neglecting temporal dependencies and directed relationships. For time series data, vector autoregressive (VAR) models have been widely used to uncover causal graph topologies \cite{granger1969investigating}. More recently, GSP methods have extended these approaches, with \cite{SegarraBlind} investigating joint identification of graph filters and input signals, and \cite{mei2016signal} proposing GSO estimation through autoregressive time-vertex system identification. \ADDED{Graph-based deep learning approaches have further advanced the field \cite{cini2023sparse, shang2021discrete}.} However, a critical limitation of these methods is that they typically operate in batch processing mode, which becomes inefficient and impractical when dealing with streaming data from complex systems where the graph topology may be time-varying \cite{timevaryinggraphs}. In such cases, batch processing would require rerunning the model each time new data arrives, leading to high computational costs.

To overcome these fundamental limitations of batch processing approaches, adaptive signal processing techniques offer a promising alternative for online learning scenarios. Algorithms like the Least Mean Square (LMS) \cite{WidrowLMS}, Recursive Least Squares (RLS) \cite{sayed2003fundamentals}, and RLS-like LMS variants \cite{rlslmsMandic} provide powerful tools for learning and tracking time-varying parameters. Recent works have introduced online schemes for estimating causal graphs from sparse VAR processes, such as TISO (Topology Identification via Sparse Online learning) and its RLS counterpart TIRSO \cite{zaman2020online}, and sparse Structural Equation Model (SEM) processes \cite{baingana2016tracking}. \ADDED{However, achieving true sparsity in online settings remains challenging, as existing methods often struggle to explicitly zero out non-causal elements despite employing sparsity-promoting regularisation \cite{Taheri2011}.} While adaptive filtering techniques have been extended to GSP \cite{LorenzoLMSGraph,YAN2022108662, LorenzoDistributed}, current research has focused primarily on signal estimation and sampling on graphs. The problem of adaptive graph topology estimation from streaming data remains largely unexplored\ADDED{, particularly sparsity-aware approaches that can reliably identify the presence or absence of causal connections.}

To address this gap, we propose \textit{AdaCGP} (\textit{\underline{Ada}ptive identification of \underline{C}ausal \underline{G}raph \underline{P}rocesses}), which leverages time-vertex GSP theory to learn and track changes in weighted graph adjacency matrices \ADDED{through a novel online sparsity-aware topology estimation method}. Our approach extends the offline approach for graph topology estimation in \cite{mei2016signal} to the online setting, enabling efficient processing of streaming data and tracking of time-varying graph topologies. Our main contributions are:
\begin{enumerate}
\item An online time-vertex adaptive filtering algorithm that extends offline approaches for graph topology estimation to streaming data scenarios, with a focus on accurate topology estimation;
\item A variable splitting approach for achieving true sparsity in online GSO estimation, enabling reliable identification of non-zero causal elements and efficient computation through sparse matrix operations;
\item A rigorous empirical analysis on synthetic data demonstrating the convergence and superiority of AdaCGP over adaptive VAR models, with prediction (forecast) error optimisation enabling reliable hyper-parameter selection and GSO sparsity estimation;
\item An application to VF recordings across increasing anti-arrhythmic drug concentrations to regulate electrophenotype complexity, demonstrating the ability of AdaCGP to capture the structure and stability of cardiac fibrillation.
\end{enumerate}

The remainder of this paper is structured as follows. Section \ref{sec:background} provides background on graph and time-vertex signal processing, followed by the formulation of our optimisation objectives in Section \ref{sec:objectives} and presentation of our adaptive algorithm in Section \ref{sec:algorithm}. We then validate our approach on synthetic data in Section \ref{sec:synthetic_data} before demonstrating its application to real-world cardiac fibrillation data in Section \ref{sec:cardiac_fibrillation}. Finally, Section \ref{sec:conclusion} presents our conclusions.

\section{Background}
\label{sec:background}
This section introduces the fundamental concepts of GSP and presents the Causal Graph Process (CGP) model for capturing temporal and structural dependencies in time-varying random graph signals.

\subsection{Graph Topology and Signals}
A graph $\mathcal{G} = (\mathcal{V}, \mathcal{E})$ is a mathematical structure that captures the relationships between a set of entities. It consists of a set of $N$ nodes, $\mathcal{V} = \{v_1, \dots, v_N\}$, and a set of edges, $\mathcal{E}$, representing their connections. A weight matrix, $\mathbf{W} \in \mathbb{R}^{N \times N}$, encodes the strength and pattern of the edge connections. This matrix, commonly referred to as the GSO in GSP, has entries $w_{ij}$ that are non-zero only if there is an edge connecting nodes $i$ and $j$, i.e., $(i,j) \in \mathcal{E}$. On this structure, a graph signal is defined as a function $f: \mathcal{V} \rightarrow \mathbb{R}$ that assigns a real value to each node, represented as a vector $\mathbf{x} = [x_1, \dots, x_N]^T \in \mathbb{R}^{N\times1}$, where $x_n$ is the signal value at node $v_n$.

\subsection{Graph Filtering}
Graph filtering is a fundamental operation in GSP that extends classical filtering to the graph domain. A graph filter is commonly expressed as a polynomial of the GSO, that is
\begin{equation}
\label{eqn:graph_filter}
H_L(\mathbf{W}, \mathbf{h}) = \sum_{l=0}^L h_l \mathbf{W}^l = h_0 \mathbf{I} + h_1 \mathbf{W} + \cdots + h_l \mathbf{W}^l,
\end{equation}
where $\mathbf{h} = [h_0, \dots, h_L]^T$ is a vector of filter coefficients and filter order $L$ determines the neighbourhood size over which it operates.

Polynomial graph filters are shift-invariant, meaning that filtering a shifted graph signal is equivalent to shifting the filtered signal, i.e., $H_L(\mathbf{W}, \mathbf{h})(\mathbf{W} \mathbf{x}) = \mathbf{W} (H_L(\mathbf{W}, \mathbf{h}) \mathbf{x})$. This shift-invariant property has two important implications. First, that the graph filter commutes with the GSO,
\begin{equation}
\label{eq:shift_invariance}
[H_L(\mathbf{W}, \mathbf{h}), \mathbf{W}] = \mathbf{0},
\end{equation}
and second, that graph filters commute with each other
\begin{equation}
\label{eq:commutativity}
[H_L(\mathbf{W}, \mathbf{h}_1), H_K(\mathbf{W}, \mathbf{h}_2)] = \mathbf{0}.
\end{equation}
Here, the commutator notation $[\cdot,\cdot]$ for two operators $\mathcal{A}$ and $\mathcal{B}$ is defined as
\begin{equation}
\label{eq:commutator}
[\mathcal{A},\mathcal{B}] \triangleq \mathcal{A}\mathcal{B} - \mathcal{B}\mathcal{A},
\end{equation}
where $[\mathcal{A},\mathcal{B}] = 0$ indicates that the operators commute \cite{mandicp2}.

\subsection{Causal Graph Process}
The CGP is a time-vertex model for random time-varying graph signals \cite{mei2016signal}. It is a special case of the more general VAR moving average time-vertex model \cite{timevertexVARMA}, which expresses the VAR coefficients as shift-invariant graph filters. The CGP model is given by
\begin{equation}
\label{eqn:cgp_model}
\mathbf{x}_t = \sum_{p=1}^{P} H_p(\mathbf{W}, \mathbf{h}_p) \mathbf{x}_{t-p} + \mathbf{w}_t,
\end{equation}
where $\mathbf{x}_t$ is the graph signal at time step $t$, $H_p(\mathbf{W}, \mathbf{h}_p)$ is an order $p$ graph filter as defined in (\ref{eqn:graph_filter}), and $\mathbf{w}_t \sim \mathcal{N}(\mathbf{0}, \mathbf{I}_{N\times N})$.

The CGP model is causal in the sense that the signal at a node at time $t$ can only be influenced by its $P$-hop neighbourhood from the previous $P$ time steps. Mathematically, this is reflected in the order of the polynomial coefficients being bounded by the lag $p$, where $p=1,\ldots, P$.

This structure models information propagation through the graph at a fixed speed (one graph shift per sampling period). While more general formulations that decouple the polynomial order from the time lag are more expressive \cite{timevertexVARMA}, the CGP model's coupled temporal and spatial dependencies provide a tractable framework for graph topology estimation \cite{mei2016signal}. The sampling frequency of discrete-time models can also be chosen to align with the propagation speed, making this assumption rather reasonable in practice.  

\begin{myrmk}
    The CGP model is inherently weakly stationary due to its formulation as a sum of shift-invariant graph filters (as defined in \cite{MarquesStationary}). While stationarity assumptions enable the definition of meaningful optimality criteria for filter design, real-world scenarios often involve non-stationary signals and time-varying graph topologies. This motivates adaptive approaches that can track gradual changes through recursive optimisation of these criteria.
\end{myrmk}

\section{Objectives for Online Optimisation}
\label{sec:objectives}
Having established the CGP model structure, we now address the problem of parameter estimation. The CGP model in (\ref{eqn:cgp_model}) can be formulated as a multivariate linear regression problem for estimating $\mathbf{W}$ and $\mathbf{h}_p$ for $p=1,\ldots, P$, where the Mean Square Error (MSE) is the optimal linear estimator \cite{edition2002probability}. In this work, we adopt the least squares method, a deterministic counterpart of the MSE estimator \cite{WidrowLMS}. Specifically, we employ the RLS-like LMS scheme \cite{rlslmsMandic}  for recursive optimisation. This approach incorporates a forgetting factor $\lambda$ into the LMS objective and has been shown to achieve fast convergence and tracking performance similar to RLS while being more numerically stable, as it does not involve matrix inversion. The least squares problem of (\ref{eqn:cgp_model}) is formulated as
\begin{equation}
\label{eq:main_objective_lsp}
\underset{\mathbf{W},\mathbf{h}}{\text{min}}\;\;\frac{1}{2}\sum\limits_{\tau=1}^t {\lambda}^{t-\tau}{\left\|\mathbf{x}_{\tau}-\sum\limits_{p=1}^PH_p(\mathbf{W}, \mathbf{h}_p)\mathbf{x}_{{\tau}-p}\right\|}_2^2,
\end{equation}
where $\mathbf{h} = [\mathbf{h}_1^T,...,\mathbf{h}_P^T]^T\in\mathbb{R}^{M\times 1}$ with $M={P(P+3)/2}$, $\mathbf{x}_\tau=\mathbf{0}$ for $\tau\leq 0$, and $\lambda\in(0,1]$. It is important to note that (\ref{eq:main_objective_lsp}) represents a non-convex optimisation problem due to the polynomial in $\mathbf{W}$. To address this, we follow the approach proposed in \cite{mei2016signal} and cast the problem into alternating steps of convex regularised least squares sub-problems.

\subsection{Solving for $\pmb{\Psi}_p=H_p(\mathbf{W}, \mathbf{h}_p)$}
The optimisation objective is first reformulated to solve for $\pmb{\Psi}_p = H_p(\mathbf{W}, \mathbf{h}_p)$ instead of $\mathbf{W}$ and $\mathbf{h}$ directly, where we denote our estimate as $\hat{\pmb{\Psi}}_p$. This transformation makes the problem quadratic in $\pmb{\Psi}_p$. Upon incorporating sparsity regularisation, we arrive at the following optimisation problem
\begin{equation}
\label{eq:subproblem1_path1}
\begin{split}
\underset{\pmb{\Psi}}{\text{min}} & \;\; \frac{1}{2}\sum\limits_{\tau=1}^t{\lambda}^{t-\tau}{\left\|\mathbf{x}_{\tau}-\sum\limits_{p=1}^P\pmb{\Psi}_p\mathbf{x}_{{\tau}-p}\right\|}_2^2 \\
& +\sum\limits_{p=1}^P\mu_p{\|\text{vec}(\pmb{\Psi}_p)\|}_1,
\end{split}
\end{equation}
where $\pmb{\Psi}=[\pmb{\Psi}_1,...,\pmb{\Psi}_P]\in\mathbb{R}^{N \times NP}$ is a concatenation of the $P$ graph filters, vec$(\cdot)$ is a vectorisation operator that stacks the columns of the matrix it acts on, and ${\|\cdot\|}_1$ is an $\ell_1$ norm, while $\mu_p$ is a constant which adjusts the degree of sparsity of the corresponding $\hat{\pmb{\Psi}}_p$.

To enforce that graph filters are shift-invariant, a soft constraint is added
\begin{equation}
\label{eq:subproblem1_path2}
\begin{split}
\;\;\underset{\pmb{\Psi}}{\text{min}}& \;\; \frac{1}{2}\sum\limits_{\tau=1}^t{\lambda}^{t-\tau}{\left\|\mathbf{x}_{\tau}-\sum\limits_{p=1}^P\pmb{\Psi}_p\mathbf{x}_{{\tau}-p}\right\|}_2^2 \\
& +\sum\limits_{p=1}^P\mu_p{\|\text{vec}(\pmb{\Psi}_p)\|}_1+\gamma\sum\limits_{i \neq j}{\|\left[\pmb{\Psi}_i,\pmb{\Psi}_j\right]\|}_F^2,
\end{split}
\end{equation}
where the additional term, weighted by $\gamma$, enforces the graph filters to commute as defined in (\ref{eq:commutativity}). This formulation results in a quartic programming problem. However, as noted in \cite{mei2016signal}, the objective is multi-convex and naturally leads to block coordinate descent as a solution, despite not being optimal in MSE. In other words, when all $\pmb{\Psi}_j$ except for $\pmb{\Psi}_i$ are held constant, the problem becomes a convex optimisation in $\pmb{\Psi}_i$.

\subsection{Estimating $\mathbf{W}$ from $\hat{\pmb{\Psi}}_1$}
From the definition of the graph filter in~\eqref{eqn:graph_filter}, $\pmb{\Psi}_1$ is a linear function of $\mathbf{W}$. Its estimate, $\hat{\pmb{\Psi}}_1$, therefore provides a reasonable (yet biased) approximation of $\mathbf{W}$. However, to determine the true $\mathbf{W}$ after obtaining $\hat{\pmb{\Psi}}$ from~\eqref{eq:subproblem1_path1} or~\eqref{eq:subproblem1_path2}, the following regularised least squares sub-problem is used
\begin{equation}
\label{eq:subproblem2}
\begin{split}
\underset{\mathbf{W}}{\text{min}} & \;\; \frac{1}{2}{\left\|\hat{\pmb{\Psi}}_1-\mathbf{W}\right\|}_2^2+\mu_1{\|\text{vec}(\mathbf{W})\|}_1 \\
& + \gamma\sum\limits_{p =2}^2{\|[\mathbf{W},\hat{\pmb{\Psi}}_p]\|}_F^2.
\end{split}
\end{equation}
where the final term enforces the shift-invariant property in~\eqref{eq:shift_invariance} by ensuring that $\hat{\mathbf{W}}$ commutes with all $\hat{\mathbf{\Psi}}_p$.

When~\eqref{eq:subproblem1_path2} is used to calculate $\hat{\pmb{\Psi}}_1$, the optimisation sub-problem in~\eqref{eq:subproblem2} may be bypassed by the approximation $\hat{\mathbf{W}}=\hat{\pmb{\Psi}}_1$, since the shift invariance property has already been enforced. We denote this latter variant of our algorithm as \textbf{Path 2}, which is detailed further in subsequent sections.

\subsection{Estimating $\mathbf{h}$}
Upon estimating $\mathbf{W}$, we can re-frame the initial objective in~\eqref{eq:main_objective_lsp} as a quadratic optimisation with respect to $\mathbf{h}$. Assuming sparsity in $\mathbf{h}$, we can substitute~\eqref{eqn:graph_filter} into~\eqref{eq:main_objective_lsp} and rearrange to give
\begin{equation}
\label{eq:subproblem3}
\underset{\mathbf{h}}{\text{min}}\;\;\frac{1}{2}\sum\limits_{\tau=1}^t{\lambda}^{t-\tau}{\left\|\mathbf{x}_{\tau}-\mathbf{Y}_{\tau}\mathbf{h}\right\|}_2^2+\eta{\|\mathbf{h}\|}_1,
\end{equation}
where $\eta$ controls for the degree of sparsity and
\begin{equation}
\label{eq:y_hoptimisation}
\mathbf{Y}_t=\left[\mathbf{x}_{t-1},\hat{\mathbf{W}}\mathbf{x}_{t-1},...,\mathbf{x}_{t-P},...,\hat{\mathbf{W}}^P\mathbf{x}_{t-P}\right].
\end{equation}
The dimensions of $\mathbf{Y}_t$ are $N \times M$, where $M=P(P+3)/2$. Despite the apparent size of $M$, practical applications often involve a much lower order, frequently satisfying $M<N$. It is important to note that this step, while informative, is not critical to our primary goal of recovering $\mathbf{W}$.

\section{Online Estimation Algorithm}
\label{sec:algorithm}
We introduce AdaCGP, an algorithm designed to achieve sparse online solutions to the objective in~\eqref{eq:main_objective_lsp}. While existing online methods such as $\ell_1$-regularised LMS \cite{chen2009sparse} and oracle algorithms \cite{LamareShrink} achieve low MSE, they rarely produce truly sparse solutions compared to offline methods like basis pursuit \cite{Taheri2011}. This is an important limitation in our context for estimating $\mathbf{W}$, since sparsity directly determines the presence or absence of edges in the graph topology, driving a causal interpretation.

To overcome this issue, we adapt the offline variable splitting approach of \cite{Schmidt2007} to the online objective in~\eqref{eq:main_objective_lsp}, reformulating the alternating $\ell_1$-regularised sub-problems in~\eqref{eq:subproblem1_path2} and~\eqref{eq:subproblem2} by splitting the target variables ($\hat{\pmb{\Psi}}$ and $\hat{\mathbf{W}}$) into their positive and negative components as
\begin{equation}
\label{eq:variable_splitting_psi}
\hat{\pmb{\Psi}}\triangleq\hat{\pmb{\Psi}}_+-\hat{\pmb{\Psi}}_-,
\end{equation}
\begin{equation}
\label{eq:variable_splitting_w}
    \hat{\mathbf{W}}\triangleq\hat{\mathbf{W}}_+-\hat{\mathbf{W}}_-,
\end{equation}
where $(\cdot)_+ \geq \mathbf{0}$ and $(\cdot)_- \geq \mathbf{0}$ contain only the positive and negative parts of $(\cdot)$, respectively. The $\ell_1$-norm can then be expressed as a product-weighted sum. For the matrices $\hat{\pmb{\Psi}}$ and $\hat{\mathbf{W}}$, this is given by
\begin{equation*}
\begin{split}
&\Vert \hat{\pmb{\Psi}} \Vert_1=\text{Tr}\left(\mathbf{1}_{N \times N}\hat{\pmb{\Psi}}_+\right)+\text{Tr}\left(\mathbf{1}_{N \times N}\hat{\pmb{\Psi}}_-\right), \\
&\Vert \hat{\mathbf{W}} \vert_1 =\text{Tr}\left(\mathbf{1}_{N \times N}\hat{\mathbf{W}}_+\right)+\text{Tr}\left(\mathbf{1}_{N \times N}\hat{\mathbf{W}}_-\right) \\
\end{split}
\end{equation*}
where $\text{Tr}(\cdot)$ is the trace operator and $\mathbf{1}_{N \times N}\in\mathbb{R}^{N \times N}$ is a matrix of unities.

These reformulations convert our first two sub-problems into non-negativity constrained optimisations \cite{Schmidt2007}, which we solve via projected stochastic gradient descent using the recursive update formulae we derive next.

\begin{myrmk}
    The variable splitting approach in (\ref{eq:variable_splitting_psi}) and (\ref{eq:variable_splitting_w}) offers potential for incorporating prior knowledge about the GSO structure. For example, when the GSO represents an adjacency matrix (elements in $\{0,1\}$), setting $\hat{\pmb{\Psi}}_- = \hat{\mathbf{W}}_- = \mathbf{0}$ naturally enforces this constraint as an inductive bias. While Laplacian matrices could be split into their positive diagonal and negative off-diagonal elements, their zero row sum proves challenging to maintain iteratively without resorting to Lagrangian methods. To maintain generality across different GSO structures, we follow the approach in \cite{mei2016signal} and study the unconstrained case.
\end{myrmk}

\subsection{Updating $\pmb{\Psi}_t$}
To minimise (\ref{eq:subproblem1_path2}) at time instant $t$, we calculate its gradient with respect to the positive part of each $p^{th}$ block of the graph filter $\hat{\pmb{\Psi}}_p$, i.e. ${(\hat{\pmb{\Psi}}_p)}_+$, which is given by
\begin{equation}
\label{eq:gradientwrtpfilter}
\begin{split}
\nabla^{(t)}_{{(\hat{\pmb{\Psi}}_p)}_+}&=\sum\limits_{\tau=1}^t{\lambda}^{t-\tau}\left(\sum\limits_{k=1}^P\hat{\pmb{\Psi}}_{k,t-1}\mathbf{x}_{{\tau}-k}\mathbf{x}^T_{{\tau}-p}-\mathbf{x}_{\tau}\mathbf{x}^T_{{\tau}-p}\right) \\
&+\mu_{p,t}\mathbf{1}_{N \times N}+\gamma\mathbf{Q}_{p,t},
\end{split}
\end{equation}
where the gradient of the commutative term is given by
\begin{equation}
\label{eq:commute_subproblem1}
\mathbf{Q}_{p,t+1}=\sum\limits_{k\neq p}^P\left(\left[\hat{\pmb{\Psi}}_{p,t},\hat{\pmb{\Psi}}_{k,t}\right]\hat{\pmb{\Psi}}_{k,t}^T-\hat{\pmb{\Psi}}_{k,t}^T\left[\hat{\pmb{\Psi}}_{p,t},\hat{\pmb{\Psi}}_{k,t}\right]\right),
\end{equation}
where we use $(\cdot)_{t}$ to denote variables $(\cdot)$ at the time instant $t$. We define $\hat{\pmb{\Psi}}_t$, $\mathbf{M}_t$, and $\mathbf{Q}_t$ as the concatenation of graph filters, sparsity and commutative terms over all $P$, given by
\begin{equation*}
\hat{\pmb{\Psi}}_t\triangleq\left[\hat{\pmb{\Psi}}_{1,t},\hat{\pmb{\Psi}}_{2,t},...,\hat{\pmb{\Psi}}_{P,t}\right]:={\hat{\pmb{\Psi}}}_{+_t}-{\hat{\pmb{\Psi}}}_{-_t},
\end{equation*}
\begin{equation*}
\mathbf{M}_t\triangleq\left[\mu_{1,t}\mathbf{1}_{N \times N},\mu_{2,t}\mathbf{1}_{N \times N},...,\mu_{P,t}\mathbf{1}_{N \times N}\right],
\end{equation*}
and
\begin{equation*}
\mathbf{Q}_t\triangleq\left[\mathbf{Q}_{1,t},\mathbf{Q}_{2,t},...,\mathbf{Q}_{P,t}\right],
\end{equation*}
where $\hat{\pmb{\Psi}}_t$ is expressed as the difference between its positive and negative parts, as in (\ref{eq:variable_splitting_psi}), and $\hat{\pmb{\Psi}}_t$, $\mathbf{M}_t$ and $\mathbf{Q}_t$ are matrices in $\mathbb{R}^{N \times NP}$.

Next, we use $\mathbf{R}_t \in \mathbb{R}^{NP \times NP}$ to denote the accumulated correlation matrix of the lagged input signals, and $\mathbf{P}_t \in \mathbb{R}^{N \times NP}$ to denote the accumulated cross-correlation matrix between the input signals $\mathbf{x}_{P,t}$ and the desired responses $\mathbf{x}_t$, where
\begin{equation*}
\mathbf{x}_{P,t}\triangleq{\left[\mathbf{x}^T_{t-1},\mathbf{x}^T_{t-2},...,\mathbf{x}^T_{t-P}\right]}^T \in \mathbb{R}^{NP\times 1}.
\end{equation*}
We can now write the recursive update formulae to $\mathbf{R}_t$ and $\mathbf{P}_t$ as
\begin{equation*}
\mathbf{R}_t\triangleq\sum\limits_{\tau=1}^t{\lambda}^{t-\tau}\mathbf{x}_{P,{\tau}}\mathbf{x}_{P,{\tau}}^T=\lambda\mathbf{R}_{t-1}+\mathbf{x}_{P,t}\mathbf{x}_{P,t}^T,
\end{equation*}
and
\begin{equation*}
\mathbf{P}_t\triangleq\sum\limits_{\tau=1}^t{\lambda}^{t-\tau}\mathbf{x}_{\tau}\mathbf{x}_{P,{\tau}}^T=\lambda\mathbf{P}_{t-1}+\mathbf{x}_t\mathbf{x}_{P,t}^T.
\end{equation*}
These expressions are then substituted into (\ref{eq:gradientwrtpfilter}) to give
\begin{equation}
\label{eq:gradient_matrix}
\mathbf{G}_t\triangleq\left[\nabla^{(t)}_{{(\hat{\pmb{\Psi}}_1)}},...,\nabla^{(t)}_{{(\hat{\pmb{\Psi}}_P)}}\right] \triangleq \hat{\pmb{\Psi}}_{t-1}\mathbf{R}_t-(\mathbf{P}_t-\gamma\mathbf{Q}_t),
\end{equation}
where it can be shown straightforwardly that $\nabla^{(t)}_{{(\hat{\pmb{\Psi}}_p)_+}} = \nabla^{(t)}_{{(\hat{\pmb{\Psi}}_p)}}$ and $\nabla^{(t)}_{{(\hat{\pmb{\Psi}}_p)_-}} = - \nabla^{(t)}_{{(\hat{\pmb{\Psi}}_p)}}$ for all $p$.

Finally, we can now express the updates to our first sub-problem as a gradient projection, that is
\begin{equation*}
{\hat{\pmb{\Psi}}}_{+_t}={\left({\hat{\pmb{\Psi}}}_{+_{t-1}}-(\mathbf{M}_t+\mathbf{G}_t)(\mathbf{A}_t\otimes\mathbf{I}_{N \times N})\right)}_+,
\end{equation*}
where $\mathbf{A}_t=\text{diag}(\alpha_1,\alpha_2,...,\alpha_P)\in\mathbb{R}^{P \times P}$ is a diagonal matrix of step-sizes for each filter block. Similarly, the update equation for ${\hat{\pmb{\Psi}}}_{-_t}$ can be obtained as
\begin{equation*}
{\hat{\pmb{\Psi}}}_{-_t}={\left({\hat{\pmb{\Psi}}}_{-_{t-1}}-(\mathbf{M}_t-\mathbf{G}_t)(\mathbf{A}_t\otimes\mathbf{I}_{N \times N})\right)}_+.
\end{equation*}
Then, finally $\hat{\pmb{\Psi}}_{t} = \hat{\pmb{\Psi}}_{+_t} - \hat{\pmb{\Psi}}_{-_t}$ can be set.

Having derived the update equations for $\hat{\pmb{\Psi}}_{t}$, we next analyse the computational complexity per iteration. The cost is dominated by the gradient calculation in (\ref{eq:gradient_matrix}). Computing $\hat{\pmb{\Psi}}_{t-1}\mathbf{R}_t$ and $\mathbf{Q}_t$ requires $\mathcal{O}(N^3P^2)$ operations in the dense case. Since our algorithm explicitly zeros out elements of $\hat{\pmb{\Psi}}_{t}$, sparse matrix operations could be used. With $S_t=\|\hat{\pmb{\Psi}}_{t}\|_0$ total non-zeros and $S_{p,t}=\max_p \|\hat{\pmb{\Psi}}_{p,t}\|_0$ maximum non-zeros per block at time $t$, the complexity reduces to $\mathcal{O}(S_{t-1}NP)$ for $\hat{\pmb{\Psi}}_{t-1}\mathbf{R}_t$ and $\mathcal{O}(S_{p,t-1}^2P)$ for $\mathbf{Q}_t$. Since $S_{t-1} \leq N^2P$, $S_{p,t-1} \leq N^2$, and $S_{t-1} \geq S_{p,t-1}$, the total complexity simplifies to $\mathcal{O}(S_{t-1}NP)$, which scales more favourably for large data applications.

\subsection{Updating $\mathbf{W}_t$}
The next step of our algorithm involves estimating the GSO, $\mathbf{W}_t$, from $\hat{\pmb{\Psi}}_t$ through either of two paths. \textbf{Path 2} provides a simple but biased estimate by setting $\hat{\mathbf{W}}_t = \hat{\pmb{\Psi}}_{1,t}$. \textbf{Path 1}, which we detail below, optimises (\ref{eq:subproblem2}) to obtain direct estimates of $\mathbf{W}_t$ at the expense of computational complexity.

In Path 1, we simplify (\ref{eq:gradient_matrix}) in the first sub-problem by setting $\mathbf{Q}_t=\mathbf{0}$ for all time steps $t$. This makes the first sub-problem convex by deferring the enforcement of shift-invariance to this phase instead. To optimise (\ref{eq:subproblem2}), the variable splitting approach from (\ref{eq:variable_splitting_w}) is followed to express $\hat{\mathbf{W}}_t$ as the difference between its positive and negative parts. Computing the gradient of (\ref{eq:subproblem2}) with respect to the positive elements of $\hat{\mathbf{W}}_t$ gives

\begin{equation}
\label{eq:gradient_subproblem2}
\mathbf{V}_t=\hat{\mathbf{W}}_{t-1}-(\hat{\pmb{\Psi}}_{1,t}-\gamma\mathbf{S}_t),
\end{equation}
where the gradient of the commutative term is given by

\begin{equation}
\label{eq:commute_subproblem2}
\mathbf{S}_t=\sum\limits_{k=2}^P\left(\left[\hat{\mathbf{W}}_{t-1},\hat{\pmb{\Psi}}_{k,t}\right]\hat{\pmb{\Psi}}_{k,t}^T-\hat{\pmb{\Psi}}_{k,t}^T\left[\hat{\mathbf{W}}_{t-1},\hat{\pmb{\Psi}}_{k,t}\right]\right).
\end{equation}

The parameter updates can again be expressed as gradient projections
\begin{equation*}
{\hat{\mathbf{W}}}_{+_t}={\left({\hat{\mathbf{W}}}_{+_{t-1}}-\beta_t(\mu_{1,t}\mathbf{1}_{N \times N}+\mathbf{V}_t)\right)}_+,
\end{equation*}
and
\begin{equation*}
{\hat{\mathbf{W}}}_{-_t}={\left({\hat{\mathbf{W}}}_{-_{t-1}}-\beta_t(\mu_{1,t}\mathbf{1}_{N \times N}-\mathbf{V}_t)\right)}_+,
\end{equation*}
where $\beta_t$ is the step-size, and $\mu_{1,t}$ is the sparsity parameter from the first sub-problem. Then, $\hat{\mathbf{W}}_t={\hat{\mathbf{W}}}_{+_t}-{\hat{\mathbf{W}}}_{-_t}$.

The computational complexity for updating $\hat{\mathbf{W}}_t$ varies between paths. While Path 2 requires $\mathcal{O}(N^2)$ operations for direct assignment of $\hat{\mathbf{W}}_t = \hat{\pmb{\Psi}}_{1,t}$, Path 1 has higher complexity of $\mathcal{O}(N^3P)$ due to the computation of the commutative term $\mathbf{S}_t$ in the dense case. However, when matrices $\hat{\mathbf{W}}$ and $\hat{\pmb{\Psi}}_k$ are sparse in (\ref{eq:commute_subproblem2}), Path 1 complexity can be reduced to $\mathcal{O}(S_{p, t}^2P)$, where $S_{p,t}$ is as defined previously.

Algorithm 1 summarises our derived adaptive algorithm for learning $\hat{\mathbf{W}}_t$. Path 1 ignores Step 9 by setting $\mathbf{Q}_t=\mathbf{0}$, executing only Step 18 to account for shift-invariance. Conversely, Path 2 includes Step 9 but skips Steps 18-22, directly assigning $\hat{\mathbf{W}}_t=\hat{\pmb{\Psi}}_{1,t}$ instead. These two paths will be tested experimentally in subsequent sections.

\begin{algorithm}
\small
\label{algo:1}
\SetKwInOut{Input}{Input}
\SetKwInOut{Output}{Output}
\SetKwRepeat{DoWhile}{do}{while}

\Input{$\mathbf{x}$, $P$}
\Output{$\hat{\pmb{\Psi}}$, $\hat{\mathbf{W}}^{\ast}$}
 Initialise $\hat{\pmb{\Psi}}_0={\hat{\pmb{\Psi}}}_{+_0}={\hat{\pmb{\Psi}}}_{-_0}=\mathbf{P}_0=\mathbf{Q}_1=\mathbf{0}$, $\hat{\mathbf{W}}_0={\hat{\mathbf{W}}}_{+_0}={\hat{\mathbf{W}}}_{-_0}=\mathbf{S}_1=\mathbf{0}$ and $\mathbf{R}_0=\mathbf{0}$\;
 $t=0$\;
 \DoWhile{$t<T^\ast$ \textnormal{(an epoch with steady-state reached)}}{
  $t=t+1$\;

  \vspace{0.75em}
  \textit{Solving for} $\hat{\pmb{\Psi}}_t$\;
  $\mathbf{x}_{P,t}={\left[\mathbf{x}^T_{t-1},\mathbf{x}^T_{t-2},...,\mathbf{x}^T_{t-P}\right]}^T$\;
  $\mathbf{R}_t=\lambda\mathbf{R}_{t-1}+\mathbf{x}_{P,t}\mathbf{x}_{P,t}^T$\;
  $\mathbf{P}_t=\lambda\mathbf{P}_{t-1}+\mathbf{x}_t\mathbf{x}_{P,t}^T$\;
  $\mathbf{Q}_t=\left[\mathbf{Q}_{1,t},\mathbf{Q}_{2,t},...,\mathbf{Q}_{P,t}\right]$ with $\mathbf{Q}_{p,t}$ according to (\ref{eq:commute_subproblem1})\;
  calculate $\pmb{\mu}_t$\;
  $\mathbf{M}_t=\left[\mu_{1,t}\mathbf{1}_{N \times N},\mu_{2,t}\mathbf{1}_{N \times N},...,\mu_{P,t}\mathbf{1}_{N \times N}\right]$\;
  $\mathbf{G}_t=\hat{\pmb{\Psi}}_{t-1}\mathbf{R}_t-(\mathbf{P}_t-\gamma\mathbf{Q}_t)$\;
  calculate $\mathbf{A}_t$\;
  ${\hat{\pmb{\Psi}}}_{+_t}={\left({\hat{\pmb{\Psi}}}_{+_{t-1}}-(\mathbf{M}_t+\mathbf{G}_t)(\mathbf{A}_t\otimes\mathbf{I}_{N \times N})\right)}_+$\;
  ${\hat{\pmb{\Psi}}}_{-_t}={\left({\hat{\pmb{\Psi}}}_{-_{t-1}}-(\mathbf{M}_t-\mathbf{G}_t)(\mathbf{A}_t\otimes\mathbf{I}_{N \times N})\right)}_+$\;
  $\hat{\pmb{\Psi}}_t={\hat{\pmb{\Psi}}}_{+_t}-{\hat{\pmb{\Psi}}}_{-_t}$\;

  \vspace{0.75em}
  \textit{Estimating} $\hat{\mathbf{W}}_t$\;
  $\mathbf{S}_t$ according to (\ref{eq:commute_subproblem2})\;
  $\mathbf{V}_t=\hat{\mathbf{W}}_{t-1}-(\hat{\pmb{\Psi}}_{1,t}-\gamma\mathbf{S}_t)$\;
  ${\hat{\mathbf{W}}}_{+_t}={\left({\hat{\mathbf{W}}}_{+_{t-1}}-\beta_t(\mu_{1,t}\mathbf{1}_{N \times N}+\mathbf{V}_t)\right)}_+$\;
  ${\hat{\mathbf{W}}}_{-_t}={\left({\hat{\mathbf{W}}}_{-_{t-1}}-\beta_t(\mu_{1,t}\mathbf{1}_{N \times N}-\mathbf{V}_t)\right)}_+$\;
  $\hat{\mathbf{W}}_t={\hat{\mathbf{W}}}_{+_t}-{\hat{\mathbf{W}}}_{-_t}$\;
\vspace{0.75em}  
}
 $\hat{\mathbf{W}}^{\ast}=\hat{\mathbf{W}}_{T^\ast}$.
\caption{Identifying the topology of $\hat{\mathbf{W}}$ ($\hat{\mathbf{W}}^{\ast}$)}
\end{algorithm}

\subsection{Estimating $\mathbf{h}_t$}
To fully identify the CGP model in (\ref{eqn:cgp_model}), the graph filter coefficients $\hat{\mathbf{h}}_t$ remain to be estimated. This phase is optional since the primary objective of this work is to estimate the GSO structure. Notice that, $\hat{\mathbf{W}}_t$ must be \textit{de-biased} before estimating $\hat{\mathbf{h}}_t$, due to the heavy regularisation involved. De-biasing is performed by fixing and optimising only the non-zero elements of $\hat{\mathbf{W}}_t$ using least squares. Note that reducing the sample size in this way can risk distorting the noise distribution from normality, potentially compromising the minimum MSE criterion for noisy or small datasets \cite{Donoho1995}.

After de-biasing $\hat{\mathbf{W}}_t$, the filter coefficients $\hat{\mathbf{h}}_t$ are estimated. Unlike zeros in the GSO, which indicate absence of edges and thus have causal significance, sparsity in $\hat{\mathbf{h}}$ serves mainly to simplify the model structure. Given the more relaxed sparsity requirements, GAR-LMS \cite{Taheri2011} is employed to arrive at the recursive update equation of $\hat{\mathbf{h}}$, given by
\begin{equation}
\label{eq:33}
\hat{\mathbf{h}}_t=\hat{\mathbf{h}}_{t-1}+\rho_t\left(\mathbf{C}_t\hat{\mathbf{h}}_{t-1}-\mathbf{u}_t+\eta_t\mathbf{b}_t\right),
\end{equation}
where
\begin{equation*}
\mathbf{C}_t=\lambda\mathbf{C}_{t-1}+\mathbf{Y}_t^T\mathbf{Y}_t,
\end{equation*}
\begin{equation*}
\mathbf{u}_t=\lambda\mathbf{u}_{t-1}+\mathbf{Y}_t^T\mathbf{x}_t,
\end{equation*}
\begin{equation*}
\mathbf{b}_t:\;b_{i,t}=\frac{\text{sign}(\hat{h}_{i,t-1})}{\epsilon+\hat{h}_{i,t-1}},
\end{equation*}
where $\mathbf{Y}_t$ contains the lagged graph shifted signals as in (\ref{eq:y_hoptimisation}), $\mathbf{C}_t \in \mathbb{R}^{M\times M}$ tracks correlations between the shifted signals, $\mathbf{u}_t \in \mathbb{R}^M$ accumulates cross-correlations with the desired response, $\mathbf{b}_t$ is the sparsity-promoting re-weighting vector, $\rho_t$ is the step-size, and $\epsilon$ prevents division by zero. This step could be further simplified by only taking the instantaneous samples into~\eqref{eq:33}, that is, $\lambda=0$, to yield
\begin{equation*}
\hat{\mathbf{h}}_t=\hat{\mathbf{h}}_{t-1}+\rho_t\left(\mathbf{Y}_t^T\mathbf{e}_t+\eta_t\mathbf{b}_t\right)
\end{equation*}
where
\begin{equation*}
\mathbf{e}_t=\mathbf{x}_t-\mathbf{Y}_t\hat{\mathbf{h}}_{t-1}.
\end{equation*}

Algorithm 2 summarises our derived adaptive algorithm for debiasing $\hat{\mathbf{W}}_t$ and estimating $\hat{\mathbf{h}}_t$. The cost of the debiasing phase is dominated by the term $\hat{\pmb{\Psi}}_{t-1}\mathbf{R}_t$, which has complexity $\mathcal{O}(N^3P^2)$ in the dense case or $\mathcal{O}(S_{t-1}NP)$ with sparse operations. The cost of updating $\hat{\mathbf{h}}_t$ is dominated by computing $\mathbf{Y}_t^T\mathbf{Y}_t$, requiring $\mathcal{O}(NP^4)$ operations for $\mathbf{Y}_t \in \mathbb{R}^{N\times M}$ with $M=P(P+3)/2$. While appearing formidable, $P$ is typically much smaller than $N$, so this update is unlikely to be the computational bottleneck in the overall algorithm.

\begin{algorithm}
\setstretch{0.5}
\small
\label{algo:2}
\SetKwInOut{Input}{Input}
\SetKwInOut{Output}{Output}
\SetKwRepeat{DoWhile}{do}{while}
\Input{$\mathbf{x}$, $P$, $\delta$}
\Output{$\hat{\mathbf{W}}$, $\hat{\mathbf{h}}$}
 All recursive variables resume from Algorithm 1\;
 $t=T^\ast$\;
 \DoWhile{$t<T$ \textnormal{(a terminal epoch)}}{
  $t=t+1$\;

  \vspace{0.75em}
  \textit{Debiasing} $\hat{\mathbf{W}}_t$\;
  $\mathbf{R}_t=\lambda\mathbf{R}_{t-1}+\mathbf{x}_{P,t}\mathbf{x}_{P,t}^T$\;
  $\mathbf{P}_t=\lambda\mathbf{P}_{t-1}+\mathbf{x}_t\mathbf{x}_{P,t}^T$\;
  $\mathbf{G}_t={\left(\hat{\pmb{\Psi}}_{t-1}\mathbf{R}_t-\mathbf{P}_t\right)}_{\hat{\mathbf{W}}}$ where $(\cdot)_{\hat{\mathbf{W}}}$ is the projection to non-zero elements of $\hat{\pmb{\Psi}}$ considering $\hat{\mathbf{W}}$\;
  calculate $\mathbf{A}_t$\;
  ${\hat{\pmb{\Psi}}}_t={\hat{\pmb{\Psi}}}_{t-1}-\mathbf{G}_t(\mathbf{A}_t\otimes\mathbf{I}_{N \times N})$\;
  Setting $\hat{\mathbf{W}}_t=\hat{\pmb{\Psi}}_{1,t}$\;

  \vspace{0.75em}
  \textit{Estimating} $\hat{\mathbf{h}}$\;
  $\mathbf{Y}_t=\left[\mathbf{x}_{t-1},\hat{\mathbf{W}}_t\mathbf{x}_{t-1},...,\mathbf{x}_{t-P},...,\hat{\mathbf{W}}_t^P\mathbf{x}_{t-P}\right]$\;
  $\mathbf{e}_t=\mathbf{x}_t-\mathbf{Y}_t\hat{\mathbf{h}}_{t-1}$\;
  $\mathbf{b}_t:\;b_{i,t}=\frac{\text{sign}(\hat{h}_{i,t})}{\sigma+\hat{h}_{i,t}}$\;
  $\hat{\mathbf{h}}_t=\hat{\mathbf{h}}_{t-1}+\rho_t\left(\mathbf{Y}_t^T\mathbf{e}_t+\eta_t\mathbf{b}_t\right)$\;
  \vspace{0.75em}
 }
 $\hat{\mathbf{W}}=\hat{\mathbf{W}}_T$, $\hat{\mathbf{h}}=\hat{\mathbf{h}}_T$
\caption{Determining the \textit{unbiased} $\hat{\mathbf{W}}$ and $\hat{\mathbf{h}}$}
\end{algorithm}

\subsection{Tuning Hyperparameters}
Our algorithm requires tuning several hyper-parameters: regularisation constants $\pmb{\mu}_t:={[\mu_{1,t},\mu_{2,t},...,\mu_{P,t}]}^T$, $\eta_t$, $\gamma$ and $\epsilon$; step-sizes $\mathbf{A}_t$, $\mathbf{\beta}_t$ and $\mathbf{\rho}_t$; and the forgetting factor $\lambda$.

Prior knowledge can guide many of these selections. For example, the $\ell_1$-norm related constants, following \cite{kim2007method}, we set
\begin{equation*}
\mu_{p,t}=\mu_p{\left\|\mathbf{P}_{p,t}-\gamma\mathbf{Q}_{p,t}\right\|}_\infty,
\end{equation*}
\begin{equation*}
\eta_t=\eta{\left\|\mathbf{Y}_t^T\mathbf{x}_t\right\|}_\infty,
\end{equation*}
where $\mathbf{P}_{p,t}$ is the $p$-th block of $\mathbf{P}_t$. Step sizes can be determined using Armijo line search \cite{boukis2010modified} or adaptive signal processing theory for stable convergence. The entries of $\pmb{\mu}$ could decrease with $p$ to reflect the decreasing sparsity of higher-order graph filters in (\ref{eqn:graph_filter}). The forgetting factor $\lambda$ should be close to 1 to balance adaptability and stability. However, parameters $\eta$ and $\gamma$ are rather unconstrained and require empirical tuning.

\subsection{Discussion on Convergence}
\ADDED{A rigorous convergence analysis for batch CGP identification using the same alternating steps of convex regularised least squares sub-problems was presented in \cite{mei2016signal}. Their analysis bounded the average excess prediction risk, defined as the difference between prediction error using the estimated and true parameters. This error scales as $O(\sqrt{\log N/K})$ for estimating the GSO, $\mathbf{W}$, and as $O(\sqrt{\log N/K^{2(\nu-\beta)}})$ for estimating the filter coefficients, $\mathbf{h}$, where $N$ is the network size, $K$ is the number of time observations, and $\nu,\beta$ are parameters used in their high-probability bounds that satisfy $0 < \beta < \nu < 1/2$.}

\ADDED{AdaCGP extends this framework through variable splitting as expressed in (\ref{eq:variable_splitting_psi}) and (\ref{eq:variable_splitting_w}), while maintaining the same core optimisation structure as \cite{mei2016signal}. The variable splitting approach reformulates the $\ell_1$-regularised problem as a non-negativity constrained optimisation \cite{Schmidt2007}, aligning with two-metric projection methods \cite{gafni1984two}. For such projection methods, \cite{gafni1984two} established that for non-critical points, there exists a step size that guarantees descent, and that every limit point of the generated sequence is a critical point. When applied to our non-negativity constraints resulting from variable splitting, these results provide theoretical support for our algorithm's convergence properties, with projected gradient theory suggesting similar convergence behaviour under standard convexity and step size conditions. The effectiveness of this technique for $\ell_1$-regularised problems has been empirically demonstrated by \cite{Schmidt2007}, showing consistent improvements in convergence speed and computational efficiency.}

\ADDED{The relationship between batch and online optimisation algorithms is well-established. For adaptive filtering, LMS converges to the Wiener-Hopf solution under small step sizes and stationarity \cite{WidrowLMS}, and RLS variants achieve faster convergence rates \cite{sayed2003fundamentals, rlslmsMandic}. Both approaches solve the same fundamental optimisation problem with different processing mechanisms: batch methods process all data simultaneously with multiple passes, while online methods process samples sequentially with a forgetting factor $\lambda$. This suggests the convergence properties established for the batch CGP algorithm \cite{mei2016signal} and variable splitting \cite{gafni1984two} should extend to our adaptive setting, given stationarity, appropriate forgetting factor, and sufficient iterations.}

\ADDED{Despite these theoretical guarantees, practical limitations remain. The theoretical framework in \cite{mei2016signal} relies on key assumptions regarding sparsity structures and stability conditions that may not hold in many real-world networks. Additionally, $\ell_1$-regularisation introduces shrinkage bias \cite{kim2007method}, and the commutativity term in \eqref{eq:subproblem1_path2} adds further complexity. Given these theoretical limitations, we complement our analysis with empirical convergence metrics, evaluating AdaCGP's ability to correctly identify the graph topology through precision, recall, and F1 scores for non-zero elements in $\mathbf{W}$. This approach recognises that identifying the presence or absence of connections is often more important than precisely estimating their weights.}

\section{Experiments on Synthetic Data}
\label{sec:synthetic_data}
To assess the performance and convergence characteristics of our algorithm variants across different graph topologies, the Normalised Mean Square Error (NMSE) was utilised as the primary metric to assess the convergence of three key aspects of our model: the prediction error from the estimated graph filter,
\begin{equation}
\text{NMSE} (\mathbf{x}_\Psi) = \frac{\Vert \mathbf{x}_t - \pmb{\hat{\Psi}}_{t-1} \mathbf{x}_{P, t} \Vert_2^2}{\Vert \mathbf{x}_t \Vert_2^2},
\label{eqn:nmse_psi}
\end{equation}
the prediction error from the estimated graph filter coefficients,
\begin{equation}
\text{NMSE} (\mathbf{x}_h) = \frac{\Vert \mathbf{x}_t - \mathbf{Y}_t\mathbf{\hat{h}}_{t-1} \Vert_2^2}{\Vert \mathbf{x}_t \Vert_2^2},
\label{eqn:nmse_h}
\end{equation}
and the error in the estimated GSO matrix,
\begin{equation}
\text{NMSE} (\mathbf{W}) = \frac{\Vert \mathbf{W} - \mathbf{\hat{W}}_t \Vert_F^2}{\Vert \mathbf{W} \Vert_F^2},
\end{equation}
where $\hat{(\cdot)}$ denotes the estimated parameters and $\Vert \cdot \Vert_F$ is the Frobenius norm. Additionally, we evaluate convergence through precision, recall, and F1 scores for the classification of non-zero elements in $\mathbf{W}$, and report the missing rate ($P_M$) and false alarm rate ($P_{FA}$) of edges once the model is in steady-state.

\subsection{Data Generation}
The AdaCGP algorithm was evaluated across four graph topologies: Random (R), Erdős-Rényi (ER) \cite{erdos1960evolution}, K-Regular (KR), and Stochastic Block Model (SBM) \cite{barabasi1999emergence}, each with $N = 50$ nodes. The random graph weights were drawn from $\mathcal{N}(0, 1)$, thresholded between $0.3w_{\max}$ and $0.7w_{\max}$ ($w_{\max}$ being the maximum absolute weight), and normalised by $1.5$ times the largest eigenvalue. For the ER graph, edges from $\mathcal{N}(0, 1)$ were thresholded between 1.6 and 1.8 in absolute value ($p_{ER} \approx 0.04$), soft thresholded by 1.5 to range between 0.1 and 0.3, and normalised by $1.5$ times the largest eigenvalue. The KR graph assigned 3 nearest-neighbour connections per node with weights from $\mathcal{U}(0.5, 1.0)$, normalised by $1.1$ times the largest eigenvalue. The SBM graph comprised $k=10$ equal-sized clusters, with intra-cluster connection probability 0.05 and inter-cluster probabilities uniform in $[0,0.04]$. Edge weights were sampled from a Laplacian distribution with rate $\lambda_e=2$, normalised by $1.1$ times the largest eigenvalue. \ADDED{Following \cite{mei2016signal}, different normalisation factors (1.1 for KR/SBM, 1.5 for Random/ER) ensure stability while preserving graph-specific properties: structured graphs require less conservative scaling than those with variable spectral properties. These factors work with filter coefficient normalisation to maintain stability in signal generation from (\ref{eqn:cgp_model}).}

Following \cite{mei2016signal}, we set model order to $P = 3$ and generated sparse filter coefficients $h_{ij}$ ($2 \leq i \leq P$, $0 \leq j \leq i$) from $2^{i+j}h_{ij} \sim 0.5\mathcal{U}(-1, -0.45) + 0.5\mathcal{U}(0.45, 1)$, normalised by 1.5 for stability. Graph signals, $\mathbf{x}_t \in \mathbb{R}^{N\times 1}$, were then generated recursively from (\ref{eqn:cgp_model}) with $\mathbf{x}_{t\leq P} = \mathbf{0}$ and $\mathbf{w}_t \sim \mathcal{N}(\mathbf{0}, \mathbf{I}_N)$. From 11,000 generated samples, we discarded the first 1000 to account for transitory effects and retained 10,000 for analysis. After hyper-parameter optimisation, this process was repeated 5 times to construct Monte Carlo estimates of our metrics and parameters.

\begin{figure}
    \centering
    \includegraphics[width=0.95\linewidth]{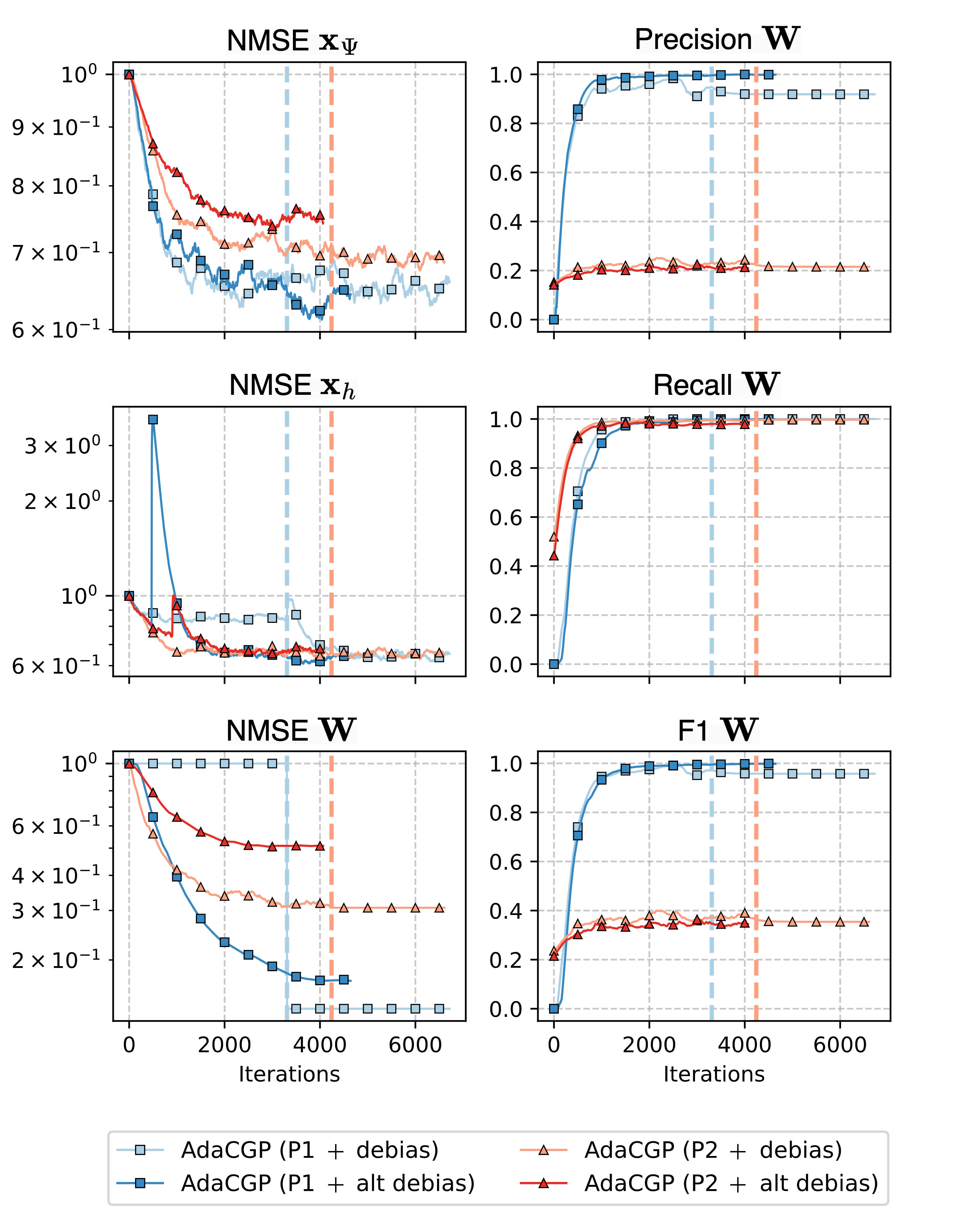}
    \caption{Convergence of AdaCGP variants on the K-Regular topology in NMSE ($\mathbf{x}_\Psi$, $\mathbf{x}_h$, $\mathbf{W}$) and the metrics for classification of non-zero elements in $\mathbf{W}$ (Precision, Recall, F1). Vertical dashed lines indicate debiasing onset.}
    \label{fig:convergence_adacgp}
\end{figure}

\subsection{Estimators: AdaCGP Variants and Baseline Models}
We evaluated four AdaCGP variants defined by two design choices: which implementation of Algorithm 1 to use (Path 1 or Path 2), and when to apply Algorithm 2 (after Algorithm 1 reaches steady-state versus in alternation at each time step). These variants were: 1) AdaCGP (P1 + debias), 2) AdaCGP (P1 + alt debias), 3) AdaCGP (P2 + debias), and 4) AdaCGP (P2 + alt debias). \ADDED{For all variants, we used simplified instantaneous updates ($\lambda=0$ in (\ref{eq:33})) for $\hat{\mathbf{h}}$ to prioritise computational efficiency, as our main goal was accurate GSO estimation rather than optimal $\text{NMSE}(\mathbf{x}_h)$ values.}

Steady-state detection was used to perform early stopping of Algorithms 1 and 2. It utilises exponential moving averages of the observable metrics $\text{NMSE}(\mathbf{x}_\Psi)$ and $\text{NMSE}(\mathbf{x}_h)$, denoted as $\sigma_t ^{(1)}$ and $\sigma_t ^{(2)}$ respectively, where $\sigma_t =  \alpha  \sigma_{t-1} + (1-\alpha) \text{NMSE}_t$ with $\alpha=0.995$. Steady-state is reached when the moving average fails to improve by $1\%$ within 500 epochs. For Algorithm 1, we monitored $\sigma_t ^{(1)}$ convergence to control the switch to Algorithm 2, as at this stage $\hat{\mathbf{h}}$ may not be meaningful due to bias in $\hat{\mathbf{W}}$. For Algorithm 2, $\sigma_t ^{(2)}$ convergence determines early stopping and defines the terminal epoch. For alternating debiasing, only $\sigma_t ^{(2)}$ is monitored.

Hyper-parameter optimisation for each variant used a random grid search over 20,000 trials, sampling elements of $\pmb{\mu}$ uniformly from (0.001, 1], $\eta$ from (0.005, 0.1] with step 0.005, $\gamma$ from (0.05, 2.0] with step 0.05, and $\lambda$ from (0.80-0.99] with step 0.01. The best parameters were selected by minimising the average of $\text{NMSE} (\mathbf{x}_h)$ during the final patience epochs of steady-state.

We benchmarked against several established baselines, both adaptive and batch methods. For adaptive methods, we included TISO and TIRSO \cite{zaman2020online}, sparse adaptive VAR models that track time-varying causality graphs online using LMS and RLS approaches, respectively. Unlike our method, these baselines do not model structural dependencies between parameters, like shift-invariance. In these cases, graph topology estimates $\hat{\mathbf{W}}_t$ are derived from the notion of VAR causality \cite{zaman2020online}, where element $(i,j)$ is causal if $\sum_{p=1}^P \mathbbm{1}(\pmb{\Psi}_{ij}^{(p)} \neq 0) > 0$, with weights assigned to causal edges as $\hat{\mathbf{W}}_{ij} = \Vert \hat{\pmb{\Psi}}_{ij} \Vert_2 \cdot \text{causal}_{ij}$. For TISO, the sparsity hyper-parameters span (0.025, 1.0] with step 0.025, while TIRSO additionally samples forgetting factors from (0.80-0.99] with step 0.01, optimising for minimal steady-state prediction error $\text{NMSE} (\mathbf{x}_\Psi)$.

\ADDED{We also included SD-SEM \cite{baingana2016tracking}, an adaptive method that tracks evolving topologies using SEMs, which model observations as linear combinations of endogenous variables with sparsity assumptions. The hyper-parameter optimisation considered forgetting factors (0.80-0.99] and sparsity penalties with values 1, 5, 10, 15, 20, and 25.}

\ADDED{For batch methods, we included the graphical LASSO (GLasso) \cite{graphLasso}, which estimates a sparse precision matrix from multivariate normal observations using the Markov property, with sparsity parameter ranging from (0.10-1.0] with step 0.10. We also evaluated GL-SigRep \cite{dong2016Learning}, which learns a graph Laplacian matrix under signal smoothness assumptions. We applied GL-SigRep per time-instant, and the hyper-parameter optimisation considered smoothness regularisation values $10^{-3}$, $10^{-2.5}$, $10^{-2}$, $10^{-1.5}$, $10^{-1}$ and sparsity strengths $10^{-2}$, $10^{-1.5}$, $10^{-1}$, $10^{-0.5}$, $10^{0}$. Finally, we included VAR and VAR+Granger \cite{granger1969investigating} models with order $P=3$, where Granger causality determines directed edges through statistical testing of VAR model coefficients.}

\ADDED{All batch methods, except GL-SigRep, were trained with all data available up to time $t$. During training, the respective objective function values were monitored for convergence, and metrics were computed on the final patience epochs of steady-state. Except for VAR+Granger, where causality tests were performed on the last 50 iterations for computational efficiency.}

To ensure optimal convergence, all adaptive autoregressive models employed an adaptive step size $\alpha_t$ (with $\mathbf{A}_t = \alpha_t \mathbf{I}_{P}$), computed as
\begin{equation}
\alpha_t = \frac{2}{\lambda_{\text{max}}(\mathbf{R}_t)} \cdot \frac{1}{\Vert \mathbf{x}_{P, t} \Vert_2^2 + \epsilon},
\label{eqn:stepsize_all}
\end{equation}
where $\epsilon$ ensures numerical stability. For SD-SEM, which is not autoregressive, the default step size of $1/\lambda_{\text{max}}$ was used. Other step-sizes $\mathbf{\beta}_t$ and $\mathbf{\rho}_t$ were calculated using the Armijo rule for automatic selection.

\begin{table*}[t]
\centering
\caption{Comparison of graph topology estimation algorithms for $N = 50$. Lower values are better for all metrics. \\Best results are in \textbf{bold}, second best are \underline{underlined}.}
\label{tab:results_table}
\setlength{\tabcolsep}{6pt}  
\setlength{\aboverulesep}{0pt}
\setlength{\belowrulesep}{0pt}
\renewcommand{\arraystretch}{1.15}
\begin{tabular}{@{}c|l|c!{\hspace{.25em}}c!{\hspace{.25em}}c!{\hspace{.25em}}c!{\hspace{.25em}}|c!{\hspace{.25em}}c!{\hspace{.25em}}c!{\hspace{.25em}}c@{}}
\toprule[1pt]\midrule[0.3pt]
&& \multicolumn{4}{c}{\textsc{Random}} & \multicolumn{4}{c}{\textsc{Erdos-Renyi}} \\
\cmidrule(lr){3-6} \cmidrule(lr){7-10}
& Algorithm & NMSE$_{\text{FC}}$ & NMSE$(\mathbf{W})$ & P$_{M}$ & P$_{FA}$ & NMSE$_{\text{FC}}$ & NMSE$(\mathbf{W})$ & P$_{M}$ & P$_{FA}$ \\
\midrule
\multirow{4}{*}{\rotatebox{90}{\textsc{Batch}}} & GLasso & $-$ & 1.72{\scriptsize$\pm$0.02} & 0.53{\scriptsize$\pm$0.01} & 0.45{\scriptsize$\pm$0.00} & $-$ & 1.97{\scriptsize$\pm$0.02} & 0.78{\scriptsize$\pm$0.01} & 0.19{\scriptsize$\pm$0.01} \\
& GL-SigRep & $-$ & 1.03{\scriptsize$\pm$0.00} & 0.01{\scriptsize$\pm$0.00} & 0.98{\scriptsize$\pm$0.00} & $-$ & 1.04{\scriptsize$\pm$0.00} & 0.01{\scriptsize$\pm$0.00} & 0.98{\scriptsize$\pm$0.00} \\
& VAR & 0.58{\scriptsize$\pm$0.01} & 2.00{\scriptsize$\pm$0.06} & 0.00{\scriptsize$\pm$0.00} & 1.00{\scriptsize$\pm$0.00} & 0.55{\scriptsize$\pm$0.01} & 1.87{\scriptsize$\pm$0.02} & 0.00{\scriptsize$\pm$0.00} & 1.00{\scriptsize$\pm$0.00} \\
& VAR + Granger & 0.56{\scriptsize$\pm$0.01} & 1.86{\scriptsize$\pm$0.02} & 0.00{\scriptsize$\pm$0.00} & 0.05{\scriptsize$\pm$0.01} & 0.56{\scriptsize$\pm$0.03} & 1.75{\scriptsize$\pm$0.00} & 0.00{\scriptsize$\pm$0.00} & 0.05{\scriptsize$\pm$0.00} \\
\midrule
\multirow{7}{*}{\rotatebox{90}{\textsc{Adaptive}}} & SD-SEM & $-$ & 1.40{\scriptsize$\pm$0.01} & 0.55{\scriptsize$\pm$0.00} & 0.44{\scriptsize$\pm$0.00} & $-$ & 1.40{\scriptsize$\pm$0.00} & 0.54{\scriptsize$\pm$0.01} & 0.42{\scriptsize$\pm$0.00} \\
& TIRSO & 0.68{\scriptsize$\pm$0.00} & 1.08{\scriptsize$\pm$0.00} & 0.00{\scriptsize$\pm$0.00} & 1.00{\scriptsize$\pm$0.00} & 0.67{\scriptsize$\pm$0.02} & 1.07{\scriptsize$\pm$0.02} & 0.00{\scriptsize$\pm$0.00} & 1.00{\scriptsize$\pm$0.00} \\
& TISO & 0.67{\scriptsize$\pm$0.01} & 1.07{\scriptsize$\pm$0.01} & 0.00{\scriptsize$\pm$0.00} & 1.00{\scriptsize$\pm$0.00} & 0.66{\scriptsize$\pm$0.03} & 1.07{\scriptsize$\pm$0.02} & 0.00{\scriptsize$\pm$0.00} & 1.00{\scriptsize$\pm$0.00} \\ \cmidrule[0.5pt]{2-10}
& P2 + alt debias & \textbf{0.53{\scriptsize$\pm$0.00}} & 0.07{\scriptsize$\pm$0.02} & 0.00{\scriptsize$\pm$0.00} & 1.00{\scriptsize$\pm$0.00} & 0.59{\scriptsize$\pm$0.01} & 0.23{\scriptsize$\pm$0.01} & 0.02{\scriptsize$\pm$0.00} & 0.05{\scriptsize$\pm$0.01} \\
& P2 + debias & \underline{0.54{\scriptsize$\pm$0.01}} & \textbf{0.04{\scriptsize$\pm$0.00}} & 0.00{\scriptsize$\pm$0.00} & 1.00{\scriptsize$\pm$0.00} & 0.58{\scriptsize$\pm$0.01} & 0.20{\scriptsize$\pm$0.02} & 0.01{\scriptsize$\pm$0.02} & 0.01{\scriptsize$\pm$0.01} \\
& P1 + alt debias & 0.56{\scriptsize$\pm$0.01} & 0.07{\scriptsize$\pm$0.00} & \underline{0.01{\scriptsize$\pm$0.01}} & \underline{0.00{\scriptsize$\pm$0.00}} & \underline{0.53{\scriptsize$\pm$0.01}} & \underline{0.03{\scriptsize$\pm$0.00}} & \underline{0.00{\scriptsize$\pm$0.00}} & \underline{0.00{\scriptsize$\pm$0.00}} \\
& P1 + debias & 0.56{\scriptsize$\pm$0.00} & \underline{0.06{\scriptsize$\pm$0.00}} & \textbf{0.00{\scriptsize$\pm$0.00}} & \textbf{0.00{\scriptsize$\pm$0.00}} & \textbf{0.52{\scriptsize$\pm$0.00}} & \textbf{0.02{\scriptsize$\pm$0.00}} & \textbf{0.00{\scriptsize$\pm$0.00}} & \textbf{0.00{\scriptsize$\pm$0.00}} \\
\midrule[0.3pt]\midrule[0.3pt]  
&& \multicolumn{4}{c}{\textsc{K-Regular}} & \multicolumn{4}{c}{\textsc{Stochastic Block Model}} \\
\cmidrule(lr){3-6} \cmidrule(lr){7-10}
& Algorithm & NMSE$_{\text{FC}}$ & NMSE$(\mathbf{W})$ & P$_{M}$ & P$_{FA}$ & NMSE$_{\text{FC}}$ & NMSE$(\mathbf{W})$ & P$_{M}$ & P$_{FA}$ \\
\midrule
\multirow{4}{*}{\rotatebox{90}{\textsc{Batch}}} & GLasso & $-$ & 3.64{\scriptsize$\pm$0.04} & 0.47{\scriptsize$\pm$0.03} & 0.19{\scriptsize$\pm$0.01} & $-$ & 14.04{\scriptsize$\pm$0.30} & 0.92{\scriptsize$\pm$0.02} & 0.03{\scriptsize$\pm$0.00} \\
& GL-SigRep & $-$ & 0.91{\scriptsize$\pm$0.00} & 0.14{\scriptsize$\pm$0.00} & 1.00{\scriptsize$\pm$0.00} & $-$ & 1.37{\scriptsize$\pm$0.00} & 0.00{\scriptsize$\pm$0.00} & 0.98{\scriptsize$\pm$0.00} \\
& VAR & 0.66{\scriptsize$\pm$0.04} & 1.14{\scriptsize$\pm$0.10} & 0.00{\scriptsize$\pm$0.00} & 1.00{\scriptsize$\pm$0.00} & 0.91{\scriptsize$\pm$0.01} & 4.57{\scriptsize$\pm$0.34} & 0.00{\scriptsize$\pm$0.00} & 1.00{\scriptsize$\pm$0.00} \\
& VAR + Granger & \underline{0.66{\scriptsize$\pm$0.02}} & 0.94{\scriptsize$\pm$0.01} & 0.00{\scriptsize$\pm$0.00} & 0.05{\scriptsize$\pm$0.01} & 0.87{\scriptsize$\pm$0.04} & 3.50{\scriptsize$\pm$0.03} & \textbf{0.08{\scriptsize$\pm$0.04}} & \textbf{0.05{\scriptsize$\pm$0.00}} \\
\midrule
\multirow{7}{*}{\rotatebox{90}{\textsc{Adaptive}}} & SD-SEM & $-$ & 1.41{\scriptsize$\pm$0.00} & 0.61{\scriptsize$\pm$0.01} & 0.43{\scriptsize$\pm$0.01} & $-$ & 3.12{\scriptsize$\pm$0.11} & 0.65{\scriptsize$\pm$0.02} & 0.31{\scriptsize$\pm$0.01} \\
& TIRSO & 0.72{\scriptsize$\pm$0.03} & 0.78{\scriptsize$\pm$0.01} & 0.00{\scriptsize$\pm$0.00} & 1.00{\scriptsize$\pm$0.00} & 0.89{\scriptsize$\pm$0.02} & 1.42{\scriptsize$\pm$0.04} & 0.00{\scriptsize$\pm$0.00} & 1.00{\scriptsize$\pm$0.00} \\
& TISO & 0.72{\scriptsize$\pm$0.03} & 0.78{\scriptsize$\pm$0.02} & 0.00{\scriptsize$\pm$0.00} & 1.00{\scriptsize$\pm$0.00} & 0.86{\scriptsize$\pm$0.02} & 1.47{\scriptsize$\pm$0.11} & 0.00{\scriptsize$\pm$0.00} & 1.00{\scriptsize$\pm$0.00} \\ \cmidrule[0.5pt]{2-10}
& P2 + alt debias & 0.76{\scriptsize$\pm$0.01} & 0.52{\scriptsize$\pm$0.02} & 0.02{\scriptsize$\pm$0.00} & 0.60{\scriptsize$\pm$0.03} & \textbf{0.78{\scriptsize$\pm$0.00}} & 0.26{\scriptsize$\pm$0.03} & 0.03{\scriptsize$\pm$0.01} & 0.91{\scriptsize$\pm$0.01} \\
& P2 + debias & 0.70{\scriptsize$\pm$0.01} & 0.33{\scriptsize$\pm$0.02} & 0.01{\scriptsize$\pm$0.00} & 0.44{\scriptsize$\pm$0.15} & \underline{0.79{\scriptsize$\pm$0.01}} & 0.19{\scriptsize$\pm$0.03} & 0.00{\scriptsize$\pm$0.04} & 0.96{\scriptsize$\pm$0.06} \\
& P1 + alt debias & 0.67{\scriptsize$\pm$0.02} & \underline{0.25{\scriptsize$\pm$0.02}} & \underline{0.02{\scriptsize$\pm$0.01}} & \underline{0.00{\scriptsize$\pm$0.00}} & 0.83{\scriptsize$\pm$0.02} & \underline{0.15{\scriptsize$\pm$0.01}} & 0.09{\scriptsize$\pm$0.02} & 0.24{\scriptsize$\pm$0.06} \\
& P1 + debias & \textbf{0.66{\scriptsize$\pm$0.01}} & \textbf{0.13{\scriptsize$\pm$0.02}} & \textbf{0.00{\scriptsize$\pm$0.00}} & \textbf{0.01{\scriptsize$\pm$0.02}} & 0.84{\scriptsize$\pm$0.01} & \textbf{0.09{\scriptsize$\pm$0.17}} & \underline{0.12{\scriptsize$\pm$0.08}} & \underline{0.08{\scriptsize$\pm$0.29}} \\
\midrule[0.3pt]\toprule[1pt]
\end{tabular}
\end{table*}

\subsection{Convergence Across Graph Topologies}
\label{sec:results_convergence}
Fig. \ref{fig:convergence_adacgp} shows the convergence of our metrics for the KR topology as a representative example. All models converge to low non-zero values for NMSE$(\mathbf{x}_\Psi)$ and NMSE$(\mathbf{x}_h)$, with Path 1 achieving lower errors than Path 2. The NMSE$(\mathbf{x}_h)$ for Path 1 showed significant improvement post-debiasing, and achieved near-zero NMSE$(\mathbf{W})$. While alternating debiasing converged to slightly higher values than steady-state debiasing, it demonstrated faster convergence in NMSE$(\mathbf{W})$ for Path 1.

The considered methods exhibited distinct differences in Precision($\mathbf{W}$), with Path 1 maintaining near-perfect precision ($\approx$1.0) compared to Path 2's plateau around 0.2. All variants achieved high Recall($\mathbf{W}$) ($\approx$1.0), though Path 2 showed faster initial increases. The F1 scores strongly favour Path 1 ($\approx$1.0 versus $\approx$0.4), with the alternating debiasing variant showing marginally better precision due to its iteration-by-iteration approach providing greater robustness to noise and misspecification of non-zero elements upon entering Algorithm 2.

As shown in Table \ref{tab:results_table}, the AdaCGP variants consistently outperformed the baselines across all graph topologies. \ADDED{To facilitate comparison across algorithms, we include $\text{NMSE}_{\text{FC}}$ in Table \ref{tab:results_table}, representing the forecast error of each model (if available), which equals $\text{NMSE}(\mathbf{x}_\mathbf{h})$ for AdaCGP variants.}

\ADDED{Specifically, AdaCGP (P1 + debias) achieves the lowest or second-lowest $\text{NMSE}(\mathbf{W})$ across all topologies, with improvements in excess of $94\%$, $98\%$, $83\%$, and $93\%$ for Random, Erdos-Renyi, K-Regular, and SBM topologies, respectively, compared to baseline models. Among adaptive algorithms, TISO and TIRSO fail to enforce sparsity ($P_{FA} = 1.00$ across all topologies), while SD-SEM sets some edges to zero but with consistently high error rates. In contrast, AdaCGP Path 1 variants maintain near-zero false alarm rates ($P_{FA} \leq 0.08$) while achieving minimal missing edge rates ($P_{M} \leq 0.12$) across all topologies. These results highlight that baseline methods struggle with their sparsity mechanisms' inability to explicitly set elements to zero, while our variable splitting approach overcomes this limitation, effectively identifying and removing irrelevant edges for an improved $P_{FA}$ and $P_{M}$ trade-off. While for batch methods, GLasso achieves lower $P_{FA}$ values but at the expense of significantly higher $P_{M}$.}

\ADDED{The SBM topology presents a particularly challenging case due to its high sparsity ($\approx1\%$ non-zero elements). Here, AdaCGP (P1 + debias) is surpassed only by VAR+Granger in $P_{M}$ ($0.12 \pm 0.08$ vs. $0.08 \pm 0.04$) and $P_{FA}$ ($0.08 \pm 0.29$ vs. $0.05 \pm 0.00$). The higher variance in $P_{FA}$ results from debiasing being applied upon convergence when facing extremely sparse topologies, where minor edge detection errors significantly impact metrics. However, despite methodological differences between VAR+Granger (sparsity obtained by offline statistical testing) and AdaCGP (sparsity obtained online), AdaCGP (P1 + debias) achieves a $\approx$$39\times$ reduction in $\text{NMSE}(\mathbf{W})$ ($0.09 \pm 0.17$ vs. $3.50 \pm 0.03$).}

\ADDED{It is important to note that $\text{NMSE}(\mathbf{x}_h)$ and $\text{NMSE}(\mathbf{W})$ evaluate different aspects of model performance (prediction accuracy vs. topology estimation accuracy). Due to the structure of our alternating optimisation framework in (\ref{eq:subproblem1_path2})-(\ref{eq:subproblem3}), minimising $\text{NMSE}(\mathbf{x}_h)$ often leads to improved $\text{NMSE}(\mathbf{W})$, while the converse is not necessarily true.}

\begin{figure*}[t]
    \centering
    \includegraphics[width=1\linewidth]{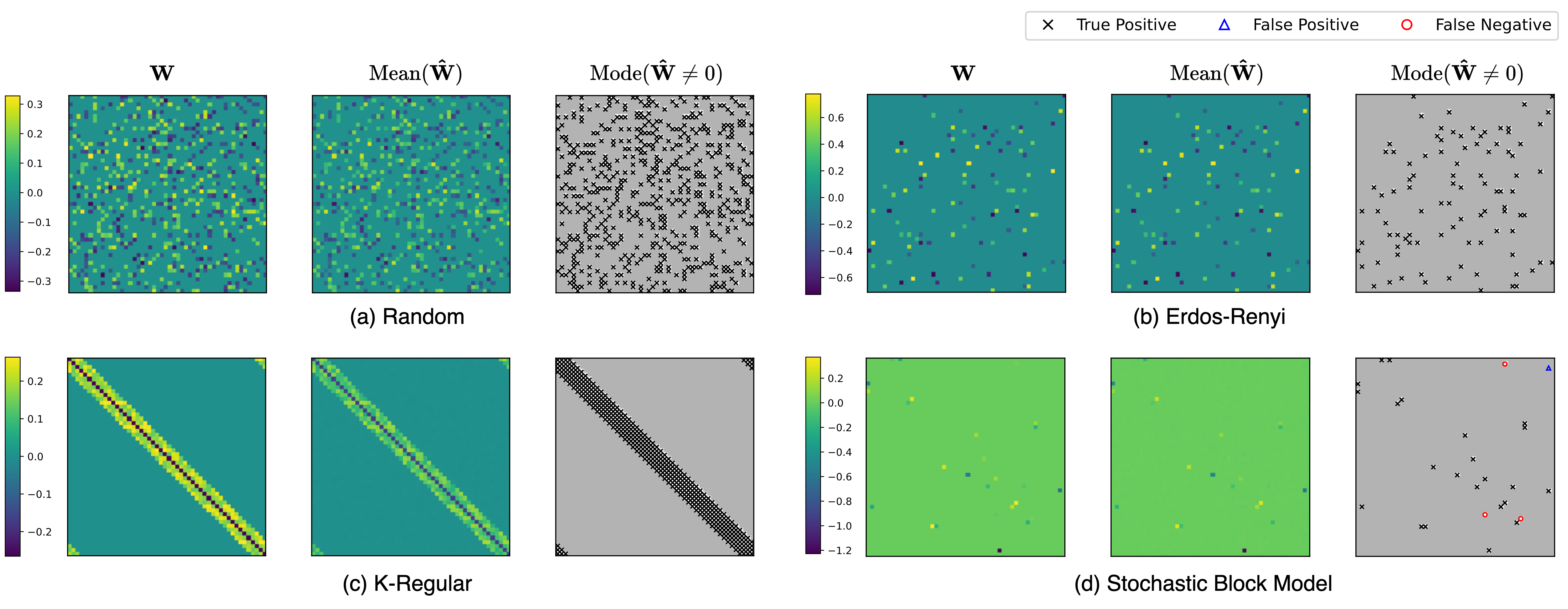}
    \caption{GSO matrices across four graph topologies: (a) Random, (b) Erdős-Rényi, (c) K-Regular, and (d) Stochastic Block Model. For each topology we show: true GSO, $\mathbf{W}$, (left); mean estimate over 20 realisations, $\text{Mean}(\mathbf{\hat{W}})$, (middle); and most commonly estimated non-zero elements, $\text{Mode}(\mathbf{\hat{W}} \neq 0)$, with highlighted true positives and false negatives (right). Results shown for the best-performing AdaCGP variant based on steady-state $\text{NMSE}(\mathbf{x}_h)$.}
    \label{fig:adjacency_estimates}
\end{figure*}

\subsection{Identification of the GSO Topology}
Fig. \ref{fig:adjacency_estimates} compares the true and estimated GSO matrices across the R, ER, KR, and SBM topologies. For each topology, we show the true GSO $\mathbf{W}$ alongside the mean estimate over 20 realisations, $\text{Mean}(\mathbf{\hat{W}})$, from the best-performing AdaCGP variant (based on lowest steady-state $\text{NMSE}(\mathbf{x}_h)$). While there is some bias in the estimated element values, particularly in the KR topology, the relative edge weights and overall structure were well preserved.

\begin{figure}
    \centering
    \includegraphics[width=\linewidth]{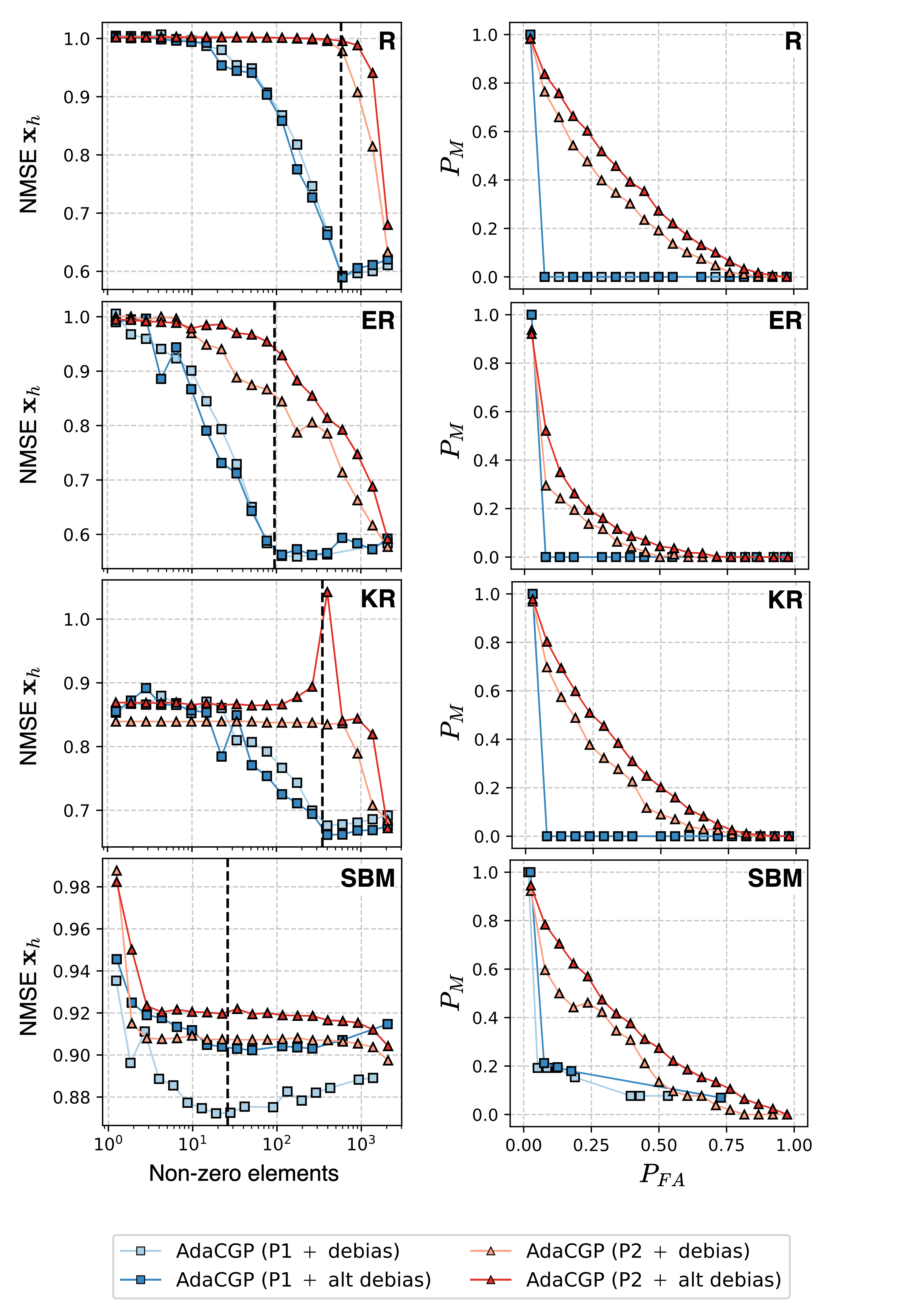}
    \caption{Performance of the AdaCGP variants for different graph topologies: R, ER, KR and SBM. Left column: $\text{NMSE}(\mathbf{x}_h)$ vs. estimated non-zeros in $\hat{\mathbf{W}}$. Vertical dashed lines indicate the true number of non-zeros. Right column: Probability of missing edges ($P_M$) vs. probability of false alarm ($P_{FA}$).}
    \label{fig:sparsity_experiment}
\end{figure}

A comparison of the most common non-zero elements over realisations, $\text{Mode}(\mathbf{\hat{W}} \neq 0)$, to the true elements in $\mathbf{W}$ reveals minimal false positives and false negatives. This demonstrates the effectiveness of AdaCGP's sparse algorithm in explicitly zeroing out GSO matrix elements while identifying most of the true edges.

\subsection{The Effect of Sparsity Regularisation on Performance}
To investigate how estimated network sparsity relates to $\text{NMSE}(\mathbf{x}_h)$, a practical metric for hyper-parameter tuning that does not require knowledge of the true network structure, we experimented with varying the sparsity regularisation strength. Using the best-performing hyper-parameters from the AdaCGP variants from before, we randomly sampled elements of $\pmb{\mu}$ uniformly from (0.001, 1] across 5,000 trials per model variant and different graph types with $N=50$ nodes.

Fig. \ref{fig:sparsity_experiment} shows the $\text{NMSE}(\mathbf{x}_h)$ versus non-zero elements in $\hat{\mathbf{W}}$ (left column) and $P_M$ versus $P_{FA}$ trade-offs (right column), where the x-axis has been uniformly binned and median results displayed. As expected, increasing sparsity regularisation reduces the estimated number of non-zero elements and increases $P_M$. Path 1 variants minimised $\text{NMSE}(\mathbf{x}_h)$ near the true sparsity level, while Path 2 variants overestimated the number of non-zero elements. Path 1 consistently achieved lower $\text{NMSE}(\mathbf{x}_h)$ and superior $P_M$-$P_{FA}$ trade-offs across all topologies, exhibiting near-perfect performance with very low $P_M$ and $P_{FA}$. These results demonstrate that optimising NMSE($\mathbf{x}_h)$ for Path 1 variants leads to accurate network sparsity estimation and thus improved causal discovery. \ADDED{We note that the results in this section serve a different purpose than those in Section~\ref{sec:results_convergence} and cannot be directly compared. While our previous analysis used optimised hyper-parameters, this section deliberately varies regularisation strength to explore the relationship between sparsity and performance metrics, with results aggregated across multiple configurations including sub-optimal ones for improved visualisation.}

\begin{figure}
    \centering
    \includegraphics[width=0.95\linewidth]{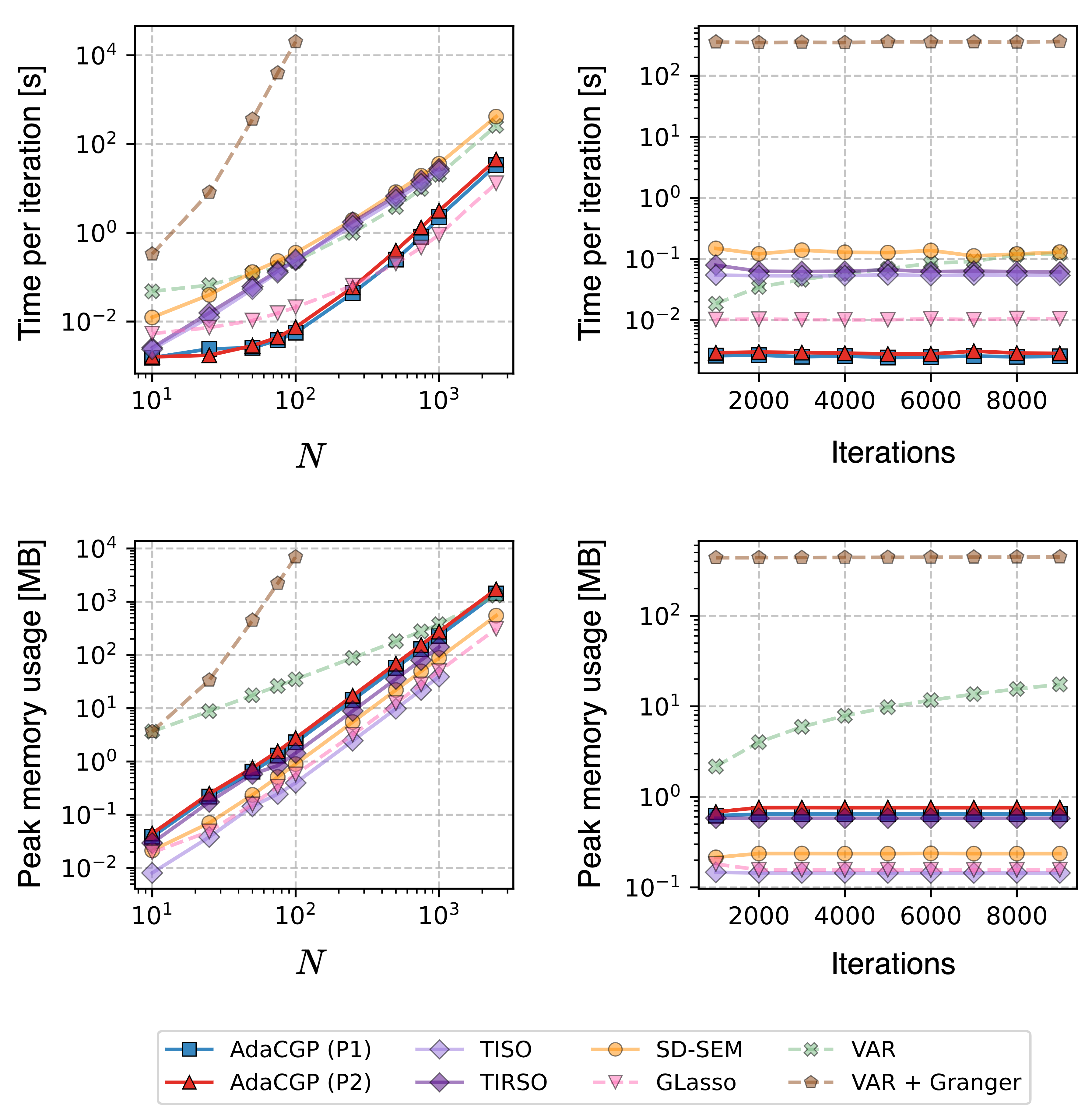}
    \caption{Computational complexity analysis of graph topology learning algorithms. Left panels: Scaling of time (top) and memory usage (bottom) with network size, $N$. Right panels: Time (top) and memory usage (bottom) as a function of iterations.}
    \label{fig:empirical_complexity}
\end{figure}

\subsection{Comparison of Runtime and Resource Utilisation}
\label{sec:complexity}
\ADDED{To assess computational efficiency, we measured CPU time per iteration and peak memory usage for the Random topology across network sizes ($N=10$ to $2500$), averaging results over three independent realisations. All algorithms were implemented in \textsc{NumPy} \cite{harris2020array} on an AMD EPYC 7513 CPU, and original source code for baselines was used when available. For all auto-regressive methods, we set $P=3$, and all adaptive algorithms employed a fixed step-size. Fig.~\ref{fig:empirical_complexity} illustrates these metrics for all algorithms, with empirical scaling exponents derived by fitting polynomial regression models to the highest 50\% of tested $N$ values. We excluded AdaCGP's debiasing step (Algorithm 2) from these experiments, since it is optional and has the same worst-case complexity as Algorithm 1.}

\ADDED{The left panels of Fig.~\ref{fig:empirical_complexity} show how the computational complexity of each algorithm scales with $N$. AdaCGP variants demonstrate empirical time complexity of $O(N^{2.91})$ for Path 1 and $O(N^{2.89})$ for Path 2, aligning with our theoretical $O(N^3)$ complexity in Section~\ref{sec:algorithm}. Despite cubic scaling, our implementation uses efficient matrix operations, achieving competitive runtimes across all tested networks ($N=10$-$2500$) and outperforming several baselines. Memory usage follows similar patterns, with AdaCGP variants maintaining $O(N^{2.00})$ scaling comparable to adaptive baselines. Table~\ref{tab:results_complexity} summarises the scaling properties for all algorithms, with VAR+Granger showing highest time complexity, $O(N^{5.81})$, due to causality testing across all potential edges. For GL-SigRep, we report the theoretical time complexity of $O(N^{2.00})$ from \cite{dong2016Learning}, as our implementation was not sufficiently optimised for large-scale benchmarking. Additionally, we observe TIRSO's time complexity of $O(N^{2.05})$ which is significantly lower than its theoretical scaling of $O(N^{3.00})$, likely because the cubic scaling has not become dominant at the network sizes tested.}

\begin{table}
\centering
\small
\caption{Scaling of per iteration time and memory complexity. Asterisk denotes theoretical scaling.}
\label{tab:results_complexity}
\setlength{\tabcolsep}{6pt}  
\setlength{\aboverulesep}{0pt}
\setlength{\belowrulesep}{0pt}
\renewcommand{\arraystretch}{1.15}
\begin{tabular}{@{}c|l|c!{\hspace{.25em}}c@{}}
\toprule[1pt]\midrule[0.3pt]
&& \multicolumn{2}{c}{\textsc{Complexity}} \\
\cmidrule(lr){3-4}
& Algorithm & Time & Memory \\
\midrule
\multirow{4}{*}{\rotatebox{90}{\textsc{Batch}}} 
& GLasso & $O(N^{2.31})$ & $O(N^{1.99})$ \\
& GL-SigRep & $O(N^{2.00})^\ast$ & $-$ \\
& VAR & $O(N^{2.42})$ & $O(N^{1.17})$ \\
& VAR + Granger & $O(N^{5.81})$ & $O(N^{3.96})$ \\
\midrule
\multirow{5}{*}{\rotatebox{90}{\textsc{Adaptive}}} 
& SD-SEM & $O(N^{2.32})$ & $O(N^{2.00})$ \\
& TIRSO & $O(N^{2.05})$ & $O(N^{2.00})$ \\
& TISO & $O(N^{1.98})$ & $O(N^{2.00})$ \\
& AdaCGP (P2) & $O(N^{2.89})$ & $O(N^{2.00})$ \\
& AdaCGP (P1) & $O(N^{2.91})$ & $O(N^{2.00})$ \\
\midrule[0.3pt]\toprule[1pt]
\end{tabular}
\end{table}

\ADDED{The right panels of Fig.~\ref{fig:empirical_complexity} demonstrate the constant per-iteration time and memory complexity of adaptive methods, a crucial property that renders them advantageous for the efficient analysis of streaming signals. In contrast, batch methods such as VAR show increasing computational demands as the training window grows (this observation is less apparent for VAR+Granger due to the significant overall cost). As discussed in Section~\ref{sec:algorithm}, sparsity could be leveraged to yield further improvements in scalability through an implementation of AdaCGP using optimised sparse matrix libraries.}

\section{Assessing the Structure and Stability of Cardiac Fibrillation}
\label{sec:cardiac_fibrillation}

While existing applications of Granger causality to fibrillation dynamics \cite{handa2020granger} use batch processing to capture average propagation patterns, our online algorithm tracks temporal changes in GSO weights and topology. This enables assessment of propagation patterns and their stability at both global and local levels, having the potential to inform clinical evaluations of the electrophenotype, where propagation appears chaotic yet maintains quantifiable structure \cite{petrutiu2006atrial, sanchez2009spectral, handa2021ventricular}.

\begin{figure*}
    \centering
    \includegraphics[width=.85\linewidth]{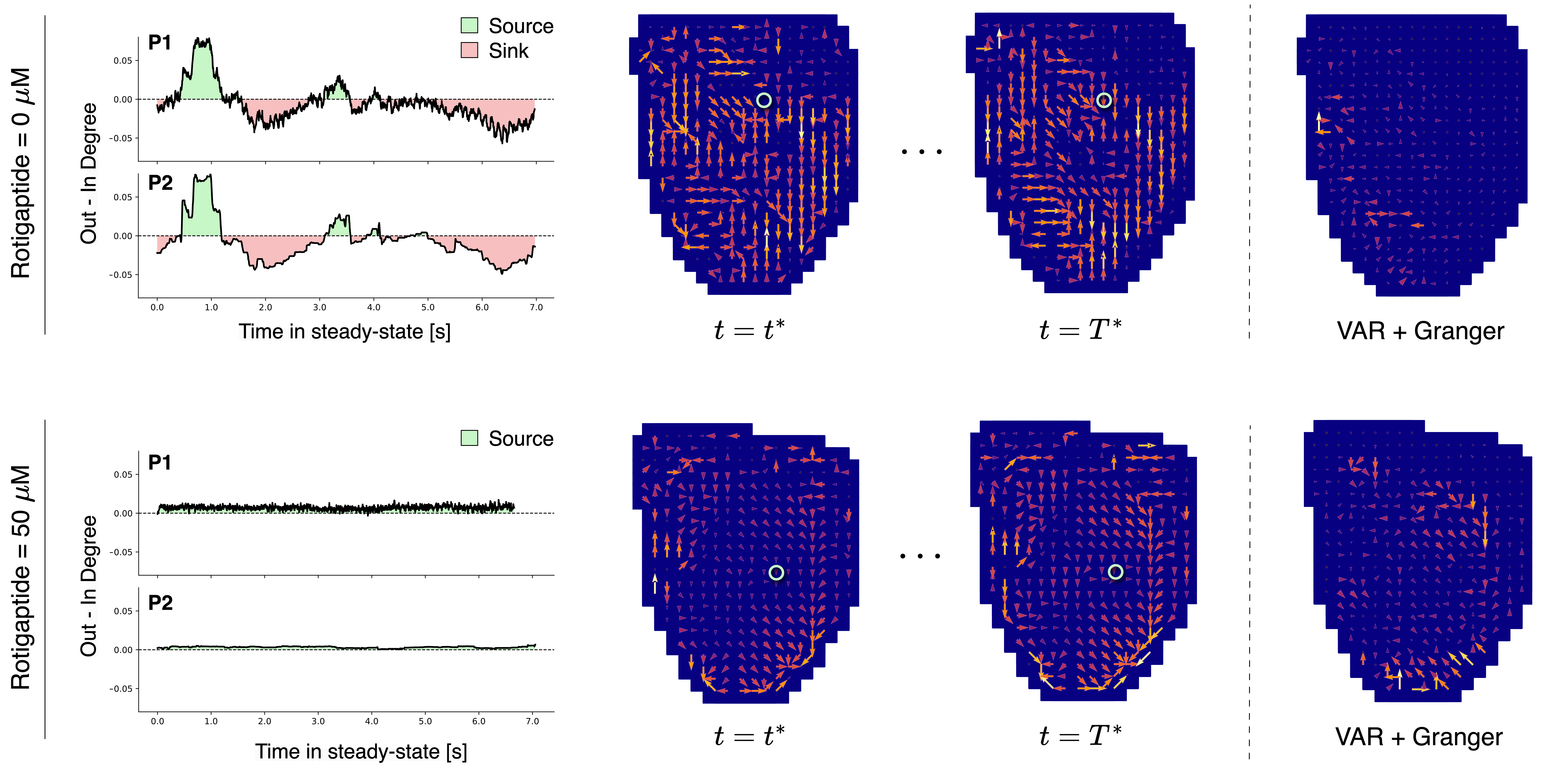}
    \caption{Local stability analysis of fibrillation dynamics. Net edge directions at steady-state start ($t=t^*$) and end ($t=T^*$) for VF (0 $\mu$M, top row) and rotigaptide 50 $\mu$M (bottom row) using AdaCGP (P1). Left panels show Out-In Degree time series for circled regions comparing P1 and P2 variants, demonstrating alternating source-sink behaviour in VF versus stable source characteristics at 50 $\mu$M. Right panels present VAR+Granger comparison. Colour and arrow length denote edge weights.}
    \label{fig:local_fibrillation_RTG}
\end{figure*}
To demonstrate these capabilities, we analysed optical mapping data of VF from an ex vivo Sprague-Dawley (SD, Charles River, Harlow, UK) rat heart during sequential administration of rotigaptide at concentrations in micromolar ($\mu$M), an anti-arrhythmic drug shown to induce more organised dynamics \cite{handa2021ventricular}. All procedures were performed in accordance with the UK Animals (Scientific Procedures) Act 1986 and ARRIVE guidelines, and were approved by the Imperial College London Ethical Review Board (project licences PEE7C76CD and PCA5EE967). Animal procedures conformed to guidelines from Directive 2010/63/EU of the European Parliament on the protection of animals used for scientific purposes. Rats were anaesthetised with 5\% isoflurane (95\% oxygen mix) and euthanised via cervical dislocation prior to heart explantation.

The experiment records fluorescence signals proportional to cardiac action potential using a 128$\times$80 pixel camera at 1000 Hz. Recordings span 10s during administration of rotigaptide in concentrations of 0, 10, 30, 50, and 80 $\mu$M, where zero represents baseline VF. \ADDED{Our preprocessing pipeline followed established standards for cardiac optical mapping data \cite{laughner2012processing, handa2021ventricular}. Briefly, raw signals were preprocessed as follows: spatial filtering ($3\times 3$ uniform kernel), temporal filtering (0-100 Hz low-pass), baseline drift removal, and amplitude normalisation. To extract phase information, we identified local minima/maxima, filtered out small amplitude fluctuations, and fitted cubic splines to generate zero-mean signals. Phase angles were then calculated from the Hilbert transform in the complex plane, capturing each region's position in its oscillatory cycle. Ventricular tissue was isolated via adaptive thresholding, with a final 30 Hz low-pass filter applied for temporal smoothing. The processed data was spatially downsampled by a factor of 4 to $\approx$350 nodes, with mean-standard deviation normalisation applied to form the resulting graph signals, $\mathbf{x}_t$.}

AdaCGP (P1 + alt debias) is deployed with gradients constrained to adjacent pixels to reflect the local connectivity of cardiac tissue. The model uses parameters $P=3$, $\gamma = 1$, $\eta = 0.01$, elements of $\pmb{\mu}$ as 0.1, and $\lambda = 0.5$, with low forgetting factor enabling rapid adaptation. \ADDED{AdaCGP (P2 + alt debias) uses identical hyper-parameters, differing only in the algorithmic approach as detailed in Section \ref{sec:algorithm}.} Step-sizes for both variants were calculated using the Armijo rule for automatic selection. \ADDED{To benchmark our approach against established methods in the cardiac electrophysiology literature, we implemented VAR+Granger as a comparative baseline, which has been widely used for functional connectivity estimation in cardiac fibrillation \cite{handa2020granger, alcaine2016multi}. Unlike our online approach, this offline method was trained on the entire time series data. Individual VAR models ($P=3$) were trained for each pixel, using only time series from adjacent pixels to reflect local connectivity and reduce computational complexity. Granger causality tests were then performed on the resulting VAR coefficients at each pixel to determine significant causal connections.}

\ADDED{Fig. \ref{fig:local_fibrillation_RTG} demonstrates local stability through edge patterns at the start ($t^*$) and end ($T^*$) of steady-state. During VF (0 $\mu$M), the magnitude and direction of net edges varied considerably between $t^*$ and $T^*$, indicating unstable propagation patterns. The highlighted region's Out-In Degree ($\sum_j \hat{W}_{ij} - \sum_i \hat{W}_{ij}$) fluctuated significantly between positive (source) and negative (sink) values, further demonstrating this instability. This dynamic behaviour is captured differently by our algorithm variants, with AdaCGP (P1 + alt de-bias) showing more rapid tracking of changes compared to AdaCGP (P2 + alt de-bias), likely due to the removal of commutativity regularisation in Path 1. In contrast, at 50 $\mu$M, edge magnitudes and directions across time points remained largely consistent, suggesting more stable propagation patterns. This stability is quantified by the highlighted region's behaviour, which maintains consistent source characteristics with minimal Out-In Degree variations that are significantly lower than during VF.}

\begin{figure*}[t]
    \centering
    \includegraphics[width=\linewidth]{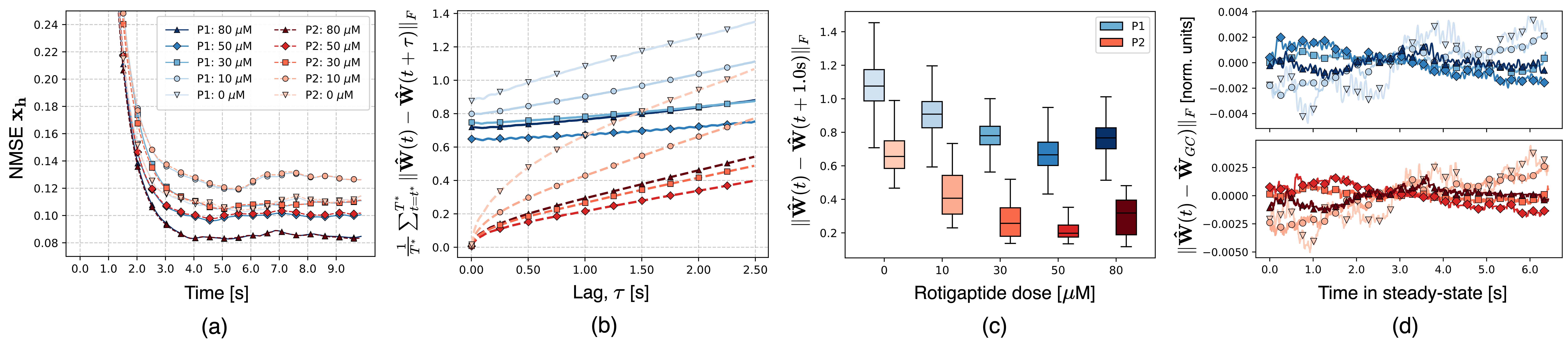}
    \caption{Global analysis of fibrillation dynamics across varying rotigaptide concentrations. (a) Prediction error $\text{NMSE}(\mathbf{x}_\mathbf{h})$ over time, showing lower steady-state errors with increasing concentrations. (b) Time-lag analysis of Frobenius norm differences between GSOs, demonstrating more stable graph topology estimates at higher concentrations and faster adaptation of AdaCGP P1. (c) Distribution of Frobenius norms between GSOs lagged by 1.0 second during steady-state, indicating more stable dynamics at higher concentrations. (d) Temporal evolution of differences between AdaCGP and VAR+Granger topologies, showing decreased variation with increasing rotigaptide concentration.}
    \label{fig:global_fibrillation_RTG}
\end{figure*}

\ADDED{When compared with VAR+Granger in Fig. \ref{fig:local_fibrillation_RTG}, the advantages of our online approach become evident. The offline nature of Granger causality, which analyses the entire time series at once, only captures average propagation patterns and misses the temporal dynamics. This limitation is particularly pronounced during baseline VF (0 $\mu$M), where Granger identifies very few causal edges due to the highly variable dynamics that get averaged out. In contrast, AdaCGP captures these dynamical variations as evidenced by the changing graph topology between $t^*$ and $T^*$. While Granger could theoretically be applied in a sliding window approach, the computational complexity of retraining for graphs of this size ($\sim$$10^2$ nodes) makes real-time analysis of cardiac data streams impractical, further highlighting the value of our adaptive approach.}

\ADDED{Fig. \ref{fig:global_fibrillation_RTG} reveals key aspects of global stability through comparisons between AdaCGP variants and VAR+Granger. First, $\text{NMSE}(\mathbf{x}_{\mathbf{h}})$ in Fig. \ref{fig:global_fibrillation_RTG}a showed lower steady-state errors with increasing rotigaptide concentrations for both AdaCGP variants, suggesting more organised dynamics consistent with the drug's anti-arrhythmic nature \cite{hsieh2016gap}. Second, Fig. \ref{fig:global_fibrillation_RTG}c showed Frobenius norms between successive GSOs exhibiting lower median values and variability at higher concentrations, indicating more stable network topologies. Our time-lag analysis in Fig. \ref{fig:global_fibrillation_RTG}b revealed important methodological differences: while both variants showed similar error patterns at larger time scales ($>$1 second), AdaCGP (P1) maintained finite error at short time lags ($<$1 second) while AdaCGP (P2) approached zero, confirming Path 1's greater sensitivity to topology changes.}

\ADDED{Notably, comparison with VAR+Granger in Fig. \ref{fig:global_fibrillation_RTG}d showed high temporal variation between AdaCGP's dynamic graphs and Granger's static graph during baseline VF (0 $\mu$M), with these variations diminishing at higher rotigaptide concentrations. This demonstrates how AdaCGP captures significant dynamic changes in graph structure that are averaged out in Granger's time-invariant representation, while confirming that both approaches align when the underlying dynamics become more stable with increased drug concentration.}

\ADDED{AdaCGP's dynamic assessment offers significant potential for clinical applications in cardiac arrhythmias. For fibrillatory conditions like AF and VF, our algorithm can identify the stability of conduction patterns maintaining the arrhythmia, directly informing treatment strategies. Specifically, by highlighting critical areas of the myocardium, AdaCGP could guide physicians' radio-frequency catheter ablation decisions regarding precise anatomical targets for intervention. This capability is particularly valuable in redo ablation cases, where previous ablation procedures have created scar tissue that distorts normal cardiac anatomy and alters myocardial electrical properties \cite{kyriakou2023right}.}

\ADDED{For successful translation to clinical practice, several challenges must be addressed. First, validating results in human subjects will require carefully designed clinical studies with appropriate ground truth comparisons. Second, while initial results show promise, further investigation is needed to systematically characterise the algorithm's performance under various noise conditions typical of clinical recordings. Third, regulatory approval pathways and developing clinically interpretable outputs will be necessary to facilitate adoption by healthcare providers. Addressing these challenges will advance clinical integration in cardiology, potentially improving arrhythmia management and enhancing ablation outcomes.}

\ADDED{Beyond cardiac applications, AdaCGP demonstrates strong potential for analysing other complex dynamic systems. In neuroscience, our approach could track dynamical functional connectivity in the brain, with applications in understanding neurological and psychiatric disorders. By identifying abnormal dynamic and transient connectivity patterns, AdaCGP could improve the characterisation of conditions such as schizophrenia, Alzheimer's disease, and depression \cite{hasan2021application, alteriis2025dysco}. Additionally, it could provide insights into how brain networks reconfigure during cognitive tasks such as memorisation, learning, and decision-making, offering a powerful tool for cognitive research.}

\section{Conclusion}
\label{sec:conclusion}
We have developed AdaCGP, an online time-vertex adaptive filtering algorithm for tracking time-varying causal graph topologies in multivariate time series. Our approach introduced a variable splitting approach for sparsity-aware adaptive filtering to efficiently identify the GSO in real-time and reliably estimate its causal elements from streaming data. Experiments across various graph topologies have demonstrated AdaCGP's superior performance over existing adaptive VAR models in both prediction accuracy and graph topology recovery. Path 1 variants have showed particularly strong performance in causal discovery while maintaining low false positive rates, with prediction error minimisation proving effective for hyper-parameter tuning and accurate sparsity estimation. Application to rat VF data revealed AdaCGP's potential for analysing complex cardiac arrhythmia. The algorithm has successfully tracked changes in cardiac organisation under varying rotigaptide concentrations, providing insights into the structure and stability of fibrillation at both global and local levels. Upon further trials in humans, this approach could help identify critical mechanisms sustaining the disorder, paving the way for personalised treatment strategies and potentially improved outcomes. Future developments could extend our framework to non-linear adaptive filtering and incorporate domain-specific GSO constraints through our variable splitting approach. Such extensions could improve the modelling of non-linear dynamical systems, such as cardiac fibrillation, where non-linear filters could better capture the chaotic dynamics and constrained GSOs, like the anisotropic Laplacian or energy-preserving shift operators \cite{scalzo2023class}, would more closely align with established physical models of cardiac conduction \cite{fentonKarma}. Overall, this work provides a principled and efficient framework for tracking evolving causal relationships in streaming data, bridging the gap between GSP theory and adaptive signal processing for real-time graph topology estimation.

\section*{Acknowledgments}
Alexander Jenkins is supported by the UKRI CDT in AI for Healthcare \url{http://ai4health.io} (Grant No. P/S023283/1). Fu Siong Ng is supported by the British Heart Foundation (RG/F/22/110078, FS/CRTF/21/24183 and RE/19/4/34215) and the National Institute for Health Research Imperial Biomedical Research Centre.

\bibliographystyle{IEEEtran}
\bibliography{bibi}

\normalfont
\vspace{-1cm}
\begin{IEEEbiography}[{\includegraphics[width=1.0in,height=1.25in,clip,keepaspectratio]{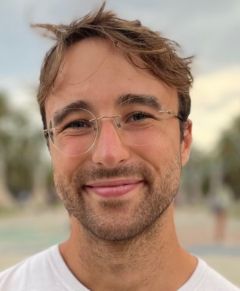}}]%
{Alexander Jenkins} is a PhD student at Imperial College London working on graph signal processing and its applications. He received his MPhys in Physics from the University of Manchester in 2019. He worked as a Research Assistant at the University of Oxford and in positions in the healthcare and financial industries.
\end{IEEEbiography}
\vspace{-1cm}
\begin{IEEEbiography}[{\includegraphics[width=1.0in,height=1.25in,clip,keepaspectratio]{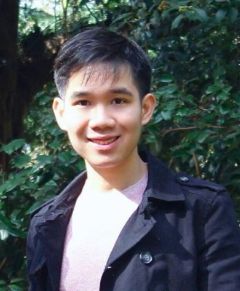}}]%
{Thiernithi Variddhisai} received his PhD in Electrical and Electronic Engineering from Imperial College London in 2020, where he worked on adaptive signal processing. He previously completed his MSc in Communications and Signal Processing, also at Imperial College London. Currently, he works in the financial industry.
\end{IEEEbiography}
\vspace{-1cm}
\begin{IEEEbiography}[{\includegraphics[width=1.0in,height=1.25in,clip,keepaspectratio]{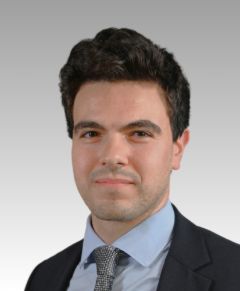}}]%
{Ahmed El-Medany} is a Cardiology specialist registrar and trainee cardiac electrophysiologist. Currently, he is a clinical research fellow and PhD student at Imperial College London working on AI in cardiology. Dr El-Medany received his medical training at Edinburgh University and subsequently an MSc in Preventive Cardiology at Imperial College London.
\end{IEEEbiography}
\vspace{-1cm}
\begin{IEEEbiography}[{\includegraphics[width=1.0in,height=1.25in,clip,keepaspectratio]{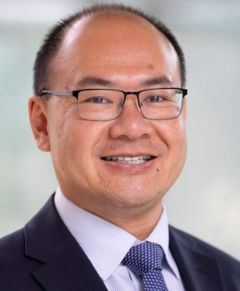}}]%
{Fu Siong Ng} is a Reader in Cardiac Electrophysiology at Imperial College London and a Consultant Cardiologist at Imperial College Healthcare NHS Trust. Dr. Ng received his medical training at St. George's University of London and his PhD from Imperial College London, where he now leads a multidisciplinary research group.
\end{IEEEbiography}
\vspace{-1cm}
\begin{IEEEbiography}[{\includegraphics[width=1.0in,height=1.25in,clip,keepaspectratio]{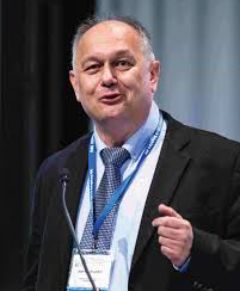}}]%
{Danilo Mandic} is a Professor of Machine Intelligence at Imperial College London. He is a Distinguished Lecturer for the IEEE Computational Intelligence Society and the IEEE Signal Processing Society, and was previously the President of the International Neural Networks Society. With over 600 publications and four research monographs, his distinguished career spans Statistical Learning Theory, Machine Intelligence, and Biomedical applications.
\end{IEEEbiography}

\vfill

\end{document}